\documentclass[pdflatex]{sn-jnl}
\usepackage[english]{babel}
\usepackage[utf8]{inputenc} 
\usepackage{csquotes}
\RequirePackage[
  backend=biber,
  style=apa6,
]{biblatex}

\usepackage{adjustbox}
\usepackage{graphicx}
\usepackage{textcomp}
\usepackage{svg}
\usepackage{booktabs}
\usepackage{xcolor}
\usepackage{mathtools}
\usepackage{subcaption}
\usepackage{caption}
\usepackage{tikz}
\usetikzlibrary{arrows.meta, positioning}
\usetikzlibrary{fit}
\usepackage{threeparttable}
\usepackage{makecell}
\usepackage{csquotes}
\usetikzlibrary{shapes.geometric, arrows}
\usepackage{soul}
\usepackage{oplotsymbl}
\usepackage{hyperref} 
\usepackage{xurl}     
\usepackage{placeins}

\newcommand{\assetNone}{\ensuremath{\circ\!\circ}}   
\newcommand{\assetU}{\ensuremath{\bullet\!\circ}}    
\newcommand{\assetC}{\ensuremath{\circ\!\bullet}}    
\newcommand{\assetUC}{\ensuremath{\bullet\!\bullet}} 

\title{A Unified Framework and Comparative Study of Decentralized Finance Derivatives Protocols}

\author[1]{\fnm{Luca} \sur{Pennella}}
\email{luca.pennella@uni.lu}

\author[2,3]{\fnm{Pietro} \sur{Saggese}}
\email{pietro.saggese@imtlucca.it}

\author[2]{\fnm{Fabio} \sur{Pinelli}}
\email{fabio.pinelli@imtlucca.it}

\author[2]{\fnm{Letterio} \sur{Galletta}}
\email{letterio.galletta@imtlucca.it}

\affil[1]{
    \orgdiv{Interdisciplinary Centre for Security, Reliability and Trust (SnT)},
    \orgname{University of Luxembourg},
    \city{Luxembourg},
    \country{Luxembourg}
}

\affil[2]{
    \orgname{IMT School for Advanced Studies Lucca},
    \city{Lucca},
    \country{Italy}
}

\affil[3]{
    \orgname{Complexity Science Hub},
    \city{Vienna},
    \country{Austria}
}

\begin{document}

\abstract{
Decentralized Finance (DeFi) applications introduce novel financial instruments replicating and extending traditional ones through blockchain-based smart contracts. 
Among these applications, DeFi derivatives protocols enable the creation and trading of decentralized derivative instruments whose value depends on underlying cryptoassets, indices, or other reference variables.
Despite their growing significance, however, they remain relatively understudied compared to other DeFi protocols, such as lending protocols and decentralized exchanges.

This paper systematically analyzes DeFi derivatives protocols, categorized into perpetuals, options, and synthetics, with the aim of comparing their instrument structures, protocol mechanisms, operational dynamics, and economic agents.
We provide a formal characterization of the main classes of decentralized derivative instruments and develop a protocol-agnostic framework that connects instrument-level specifications, market-state variables, and protocol-level mechanisms.
We complement the analytical framework with numerical simulations that evaluate how derivative positions evolve under varying economic conditions, including changes in underlying asset prices, volatility, protocol-specific fees, and leverage.

Overall, this study provides a structured analytical framework for understanding and comparing the design and functioning of decentralized finance derivatives protocols.
\\[1ex]

}


\maketitle

\begin{center}
\small
\textit{A peer-reviewed version of this article has been published in 
Electronic Markets. Please cite the published version: 
\url{https://doi.org/10.1007/s12525-026-00925-9}.}
\end{center}

\vspace{0.5em}


\section{Introduction}
\label{sec:intro}

Derivatives protocols are an established category of Decentralized Finance (DeFi) applications that enable users to create, trade, or manage derivative instruments. These financial instruments are implemented through smart contracts deployed on a Distributed Ledger Technology (DLT), such as Ethereum. 
Their value is derived from an underlying asset or index, such as cryptocurrencies, fiat currencies, commodities, or other financial indicators.

\begin{figure}[t]
  \centering
  \includegraphics[width=0.75\textwidth]{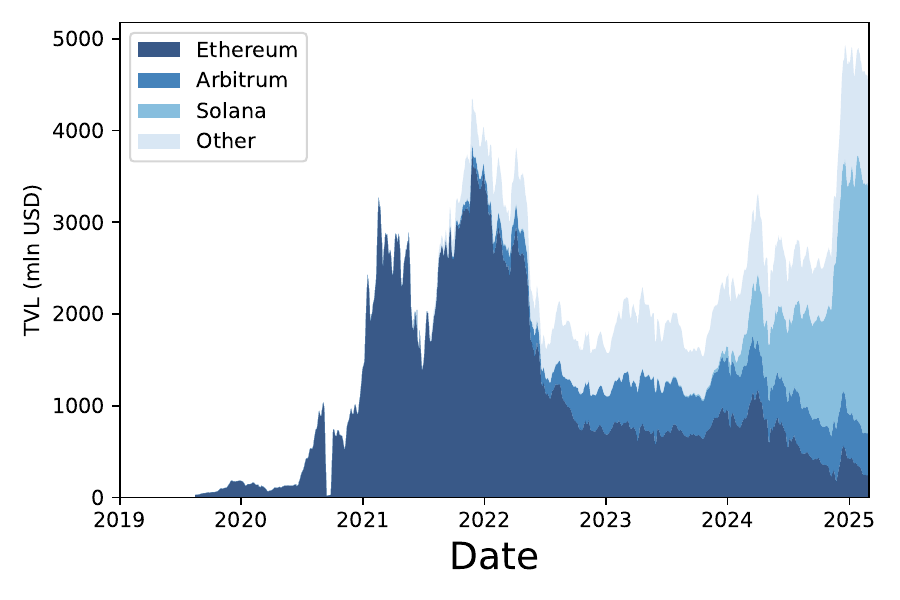}
  \caption{Total Value Locked (TVL) evolution over time for the derivative protocols reported in Table~\ref{tab:protocols}, distinguished by chain. Data extracted from DeFiLlama (\citeyear{defillama}).}
  \label{fig:TVL_all}
\end{figure}

Despite a strong initial interest within the DeFi community, the growth of derivatives protocols experienced a contraction after mid 2022. 
However, as Figure~\ref{fig:TVL_all} shows, the total U.S. dollar value of digital assets deposited in the smart contracts of major derivatives protocols - known as the \emph{Total Value Locked} (TVL) - has started increasing again rapidly since 2024, driven mainly by new blockchain platforms, such as Solana and Arbitrum. 
This revived growth appears to be associated with the emergence of new market entrants that are challenging existing incumbent decentralized applications implemented on the Ethereum blockchain.
This scenario suggests an evolving, competitive landscape within the DeFi derivatives sector and a renewed interest from developers and DeFi users.

Derivatives protocols are commonly mentioned as one of DeFi's core decentralized applications (dApps), along with Protocols for Lending Funds (PLFs), Decentralized Exchanges (DEXs) with Automated Market Makers (AMMs), Yield Aggregators, and Liquid Staking protocols~\parencite{schar:decentralized:2021,werner2022sok,harvey2021defi,xu2022reap, kraner_et_al:LIPIcs.AFT.2025.9}. 
Like other DeFi dApps, derivatives protocols rely on blockchain technology and web services to offer decentralized financial instruments. More broadly, they may contribute to reshaping financial and web-based ecosystems through algorithmic automation, competitive financial engineering, and new forms of openness in the provision of financial services~\parencite{auer2024technology}.
In contrast to other well-researched DeFi applications, however, they remain relatively understudied.

Indeed, the existing work on derivatives in the crypto ecosystem focuses mainly on tokens~\parencite{luo2024piercing,gudgeon2020defi} or on products traded on centralized exchanges~\parencite{he2022fundamentals}. 
Existing research has not yet provided a systematic analytical framework for decentralized derivatives protocols and the services they offer.


\begin{table}[t]
  \centering
  \scriptsize
\caption{List of protocols analyzed, selected as the top five protocols by TVL on DeFiLlama within the Derivatives, Synthetics, and Options categories. The Derivatives category primarily includes protocols issuing perpetuals. The table reports the protocol category, main chain, TVL in US\$ on 31 December 2024, and the TVL percentage variation during 2024.}
  \label{tab:protocols}
  \begin{tabular*}{\textwidth}{@{\extracolsep{\fill}}llcrr}
    \toprule
    \textbf{Protocol} & \textbf{\makecell[l]{Category}} 
      & \textbf{\makecell[c]{Main \\ Chain}} & \textbf{\makecell{TVL \\ (m USD)}} & \textbf{\makecell{TVL Change \\ (2024)}} \\
    \midrule
    \makecell[l]{Jupiter} & Perpetuals & Solana      & 1720.88 & +1864.36\% \\
    Drift Trade           & Perpetuals & Solana      & 802.55  & +595.48\%  \\
    GMX                   & Perpetuals & Arbitrum    & 696.06  & -29.07\%   \\
    dYdX                  & Perpetuals & dYdX        & 431.56  & +23.69\%   \\
    Hyperliquid           & Perpetuals & Hyperliquid & 398.65  & +151.84\%  \\
    Synthetix             & Synthetics & Ethereum    & 381.33  & -50.23\%   \\
    Derive                & Options    & Ethereum    & 82.44   & +2202.17\% \\
    Alchemix              & Synthetics & Ethereum    & 67.79   & -7.15\%    \\
    Youves                & Synthetics & Tezos       & 47.60   & +10.02\%   \\
    GammaSwap             & Options    & Arbitrum    & 27.36   & +34860.89\% \\
    Deri V4               & Options    & Linea       & 24.67   & +96.20\%   \\
    Metronome             & Synthetics & Ethereum    & 15.88   & +1.97\%    \\
    Hegic                 & Options    & Arbitrum    & 13.40   & -10.65\%   \\
    SOFA.org              & Options    & Arbitrum    & 6.94    & +1230.15\% \\
    Taiga                 & Synthetics & Acala       & 2.76    & -85.64\%   \\
    \bottomrule
  \end{tabular*}
\end{table}

Our paper addresses this research gap by systematically reviewing existing DeFi derivatives protocols. We examine their main design features, the instrument classes they support, and the similarities and differences in how these protocols are implemented in practice.

Building on this protocol analysis, we develop a two-layer framework for DeFi derivatives. The first layer formalizes the economic structure of the main derivative instrument classes observed in decentralized markets, namely perpetuals, options, and synthetic assets. The second layer maps the protocol components and operational mechanisms through which these instruments are instantiated, priced, collateralized, maintained, and settled on-chain.

More specifically, our framework includes: (i) a formal representation of the derivative instrument classes considered; (ii) a taxonomy of the main economic agents and protocol components involved in their functioning; (iii) a characterization of the main interaction mechanisms and operational dynamics through which these instruments are issued, traded, collateralized, and settled; (iv) a systematization of how different components are implemented across protocols in practice; and (v) a simulation environment that operationalizes selected elements of the formal representation to explore the interaction between market conditions and protocol design variables under stylized assumptions.

To develop this framework, we follow a structured process consisting of protocol selection, source analysis, and framework derivation. We select a sample of major derivatives protocols, reported in Table~\ref{tab:protocols}, analyze their documentation, interfaces, smart contracts, and technical sources, and extract their main design and operational characteristics.\footnote{The entire list of sources used is reported in Table~\ref{tab:resr} in the Appendix.} Through this approach, we compare heterogeneous DeFi implementations within a common analytical structure, providing a structured basis for understanding the design and functioning of derivatives protocols in decentralized finance.

The remainder of the paper is structured as follows. Section~\ref{sec:background} provides background on derivatives in traditional finance and their evolution within the crypto sector.
In Section~\ref{RW}, we review the related work. 
Section~\ref{sec:methodology} presents the methodology adopted to select the protocols analyzed and the data sources used. 
Section~\ref{sec:concept} introduces the analytical formalization of the main derivative instrument classes. Section~\ref{sec:protocols} then extends this perspective to the protocol layer by presenting the key actors, components, and operational dynamics of decentralized derivatives protocols.
In Section~\ref{math_function} we complement the theoretical framework with Monte Carlo simulations to analyze how underlying price dynamics and protocol design variables affect position outcomes under varying conditions.
Finally, Section~\ref{sec:conclusions} concludes and outlines directions for future research. The appendix provides additional tables, the full list of sources, and further background details on derivatives in traditional finance.

\section{Background}
\label{sec:background}

In this section, we provide background information to position derivatives protocols into the landscape of DeFi and relate them to their corresponding equivalents in traditional finance.

\subsection{Derivative contracts in traditional finance (TradFi)}

In TradFi, a derivative contract is a financial instrument whose value is derived from an underlying variable, which can range from the price of a traded asset to the performance of an index, underlying commodity or currency rate, among others. It represents an agreement between two parties to exchange value based on changes in the underlying variable's valuation over time.

The global derivatives market is massive and complex~\parencite{release2023otc}, thus a comprehensive discussion of all existing derivatives contracts goes beyond the goals of this study. 
In the following, we introduce key elements and the primary instruments traded in traditional finance.

\emph{Forward contracts} are customized contracts between two parties, typically traded over-the-counter (OTC), to buy or sell an asset at a specified future date for a price previously agreed upon. 
OTC refers to a kind of market where transactions occur bilaterally between participants, without the involvement of a centralized exchange. 
\emph{Futures contracts} are similar to forwards, but they are standardized and traded on exchanges. They are typically marked to market on a daily basis and converge to the spot price of the underlying asset as maturity approaches.
\emph{Options} grant the buyer the right, but not the obligation, to buy (call) or sell (put) an asset at a specified price within a defined period. This advantage comes at the cost of an up-front fee. 
\emph{Swaps} are agreements between two parties to exchange cash flows or other financial instruments. The most prominent ones are interest rate swaps, credit default swaps, currency and equity swaps.

Derivatives are typically traded on centralized exchanges or OTC.
While the two mechanisms exhibit differences, in both cases several intermediaries are involved in the contract creation, from brokers, who facilitate user access to exchanges and support negotiations, to financial institutions, which create and manage derivative trading, and clearinghouses, which act as central counterparties to guarantee settlement and mitigate risk.
More details on these products are in Appendix~\ref{sec:app}.

\subsection{Derivative contracts in crypto centralized finance (CeFi)}

Blockchain related derivatives can refer either to conventional derivative contracts written on cryptoassets or to instruments whose issuance, trading, or settlement relies directly on blockchain infrastructure.
The former are financial instruments typically created and traded through centralized cryptocurrency exchanges (CEXs) such as Binance or BitMex.\footnote{\url{https://www.binance.com/en} and \url{https://www.bitmex.com/}}
These trading platforms enable customers to trade cryptocurrencies or derivatives built on top of them in a centralized ecosystem, with trades executed on a platform rather than on a blockchain and the exchange acting as a custodian and counterparty. 
CEXs and other cryptoasset service providers are part of Centralized Finance (CeFi), an ecosystem that mirrors traditional financial systems despite being based on distributed ledger technologies~\parencite{saggese2024assessing}.

Whilst derivatives offered by CEXs share similarities with TradFi products, the CeFi derivatives market has distinct characteristics. 
One primary difference is that the derivatives market in CeFi is largely dominated by perpetuals, a type of contract similar to futures but without an expiration date. 
Interestingly, perpetuals were first proposed by \textcite{shiller1993measuring} with no relation to cryptocurrency markets. However, they were not implemented in practice for a long time, until they started gaining significant traction and wider adoption in cryptocurrency markets, when they were pioneered by BitMEX~(\citeyear{perpetual:bitmex}).
Unlike standard futures, perpetuals rely on a funding-fee-based mechanism in which traders periodically exchange payments to keep the contract price aligned with the underlying reference price.

\subsection{Derivative contracts in decentralized finance (DeFi)}

A distinct class of blockchain based derivatives consists of instruments natively implemented on blockchain infrastructure.
These are financial products created by Derivatives protocols, i.e., DeFi applications that leverage DLTs to replicate existing financial instruments in a decentralized system.
Unlike derivatives offered in centralized ecosystems, these financial instruments are issued and traded automatically and deterministically on-chain through smart contracts. In many implementations, users interact with smart contracts or protocol-managed liquidity pools rather than directly with a bilateral counterparty, although other designs rely on order books or market makers~\parencite{auer2024technology}. Trading and clearing mechanisms are meant to be implemented on-chain as well, aiming to eliminate reliance on centralized intermediaries.

According to leading DeFi data aggregators, derivative protocols are commonly grouped into three categories: perpetuals (often listed under ‘derivatives’), synthetics, and options (see Table~\ref{tab:protocols}).
Perpetuals are similar to those implemented in CEXs such as BitMex; synthetic assets are contracts that track the performance of various assets without requiring their ownership; and option-based instruments replicate the economic logic of their traditional finance counterparts.

DeFi derivatives typically reference crypto-native assets as the underlying asset, but may also track traditional financial indices or synthetic representations of real-world assets.
Notably, DeFi derivatives may thus rely on oracles, i.e., blockchain-based services that feed external (off-chain) data to a blockchain or smart contract, to correctly price derivatives~\parencite{eskandari2021sok}.

Note that issuing and trading derivatives on cryptocurrencies takes place mostly on centralized cryptocurrency exchanges at the moment, with daily trading volume in the order of magnitude of billions~\parencite{ackerer2024perpetual}.  However, although we acknowledge that the CeFi market is larger than that of DeFi today, the former does not differ substantially from traditional derivatives, while the latter represents a rapidly growing segment of the crypto market and a significant innovation in financial infrastructure design. For this reason, we will focus on this category in the remainder of the paper.

\section{Related Work} \label{RW}

Derivative protocols are often cited as one of the most relevant decentralized applications in DeFi~\parencite{schar:decentralized:2021,werner2022sok,auer2024technology}. 
Despite this, they remain relatively understudied compared to both cryptocurrency-based derivatives traded on CEXs and other DeFi applications, and to the best of our knowledge, a unified framework abstracting their design space is missing.

A large body of literature has examined perpetual derivatives and other instruments issued by centralized exchanges~\parencite{soska2021towards,rahimian2024shortfall}, focusing e.g. on pricing dynamics~\parencite{alexander2020bitmex,konczal2025pricing,brini2024pricing}, hedging strategies~\parencite{alexander2023hedging,cheng2021liquidation}, market liquidity and quality~\parencite{ruan2022perpetual}, contract design and market microstructure~\parencite{de2022arbitrage}, and deriving no-arbitrage pricing for perpetual futures under frictionless assumptions~\parencite{he2022fundamentals, ackerer2024perpetual}.

Similarly, the academic community has extensively studied DeFi applications like protocols for loanable funds~\parencite{bartoletti:sok:2021,gudgeon2020defi} and DEXs with AMMs~\parencite{lehar2025decentralized,bartoletti2022theory,xu:sok:2023}, which respectively enable users to lend (borrow) and trade crypto assets. Moreover, a systematization of prior work abstracting the main design choices and mechanisms of other DeFi applications like yield aggregators~\parencite{cousaert2022sok,xu2022reap} and staking protocols~\parencite{gogol2024sok} already exists.

In contrast, the academic literature on DeFi derivative protocols is limited and fragmented.
%
One stream of research focuses on perpetuals.
\textcite{angeris2023primer} formalize perpetual futures in a parameterized, mathematically rigorous manner, and \textcite{do2024novel} introduce an AMM-based mechanism for decentralized perpetual futures, aiming to reduce liquidation and bankruptcy risks respectively for traders and liquidity providers.
Closer to our aims, \textcite{chen2024perpetual} systematize differences between CeFi and DeFi perpetuals, highlighting how varying design mechanisms shape market dynamics and traders' behavior.
A second stream of literature focuses on the DeFi options and synthetics market. 
\textcite{andolfatto:decentralized:2024} and \textcite{singh2024option} compare CeFi and DeFi option markets. The former argue that higher costs in DeFi reflect thinner liquidity and limited retail demand, while the latter examine the main design choices of these markets and identify key technical, financial, and adoption-related challenges.
\textcite{zhang2024practical} propose UP-BLOC, an options contract protocol that is applicable to any blockchain and allows the buyer to engage in trading without locking assets.
Finally,~\textcite{rahman2022systematization} conduct an overview of synthetic assets and a preliminary analysis of their basic mechanisms.

In summary, existing studies on DeFi derivatives protocols focus on individual derivative protocol categories or investigate a limited set of features, rather than comparing their full design space across protocols. By contrast, we adopt a broader perspective, analyzing these protocols --- perpetuals, options (expiring and everlasting), and synthetics --- altogether, and comparing both their financial elements and protocol dynamics within a unified formal approach.
To the best of our knowledge, no prior work (i) builds such a unified cross-protocol conceptualization; (ii) maps actors, components, and operational flows across protocol designs; and (iii) complements it with a simulation framework that quantifies how protocol parameters and market conditions affect profitability and liquidation risk of a position; thus differentiating our work from instrument- or protocol-specific studies.
To this end, we develop a structured approach, introduced in the following section, to support our cross-protocol analysis and formalization.

\section{Methodology and Data}\label{sec:methodology}

This section describes the methodology used to select the derivatives protocols included in our systematization and the data sources analyzed to derive the proposed framework.
We follow a structured process that consists of three main phases: protocol selection, data source analysis, and derivation of the formal framework. 
In these phases, we combine a data-driven approach for protocol selection with qualitative and technical analysis to extract relevant information.
Below, we detail each of them.

\paragraph{Protocol selection}
To identify the most prominent derivatives protocols, we rely on DeFiLlama~(\citeyear{defillama}), a widely used DeFi data aggregator that tracks metrics like TVL, publishes protocol performance indicators, and categorizes DeFi protocols according to the financial services they offer. In addition to the \textit{Derivatives} category, which comprises primarily protocols issuing perpetuals, DeFiLlama includes other related categories, such as Synthetics, Options, Risk Curators, and Insurance.\footnote{A full list of DeFiLlama categories can be found at \url{https://defillama.com/categories}} The latter two mostly focus on coverage against on-chain risks, and we consider them to be beyond the scope of our study. We therefore focus on the three subcategories \emph{Perpetuals}, \emph{Synthetics}, and \emph{Options}. 
For each category, we selected the top five protocols based on TVL
serving as a proxy for user adoption and capital engagement.\footnote{While we acknowledge that TVL is an imperfect metric currently lacking a standardized methodology~\parencite{luo2024piercing,saggese2025towards}, it is widely accepted as a measure of performance and scale of DeFi protocols and DeFi more broadly.}
The rationale behind our selection is that, given the highly dynamic nature of DeFi, covering all derivatives protocols would be challenging; therefore, we focus on the top ones, which are the most widely adopted and assume that their mechanisms and dynamics are representative of the largest part of the ecosystem.

Once we identified a list of candidate protocols, we performed a filtering step as follows.
We manually verified the accessibility of each protocol’s public interface, including the website, the trading application (dApp), and the documentation.
Protocols lacking accessible interfaces or documentation were excluded.
For instance, Outcome Finance~(\citeyear{outcomefinance_defillama}) was excluded due to inaccessibility, while Opyn~(\citeyear{opyn})  was omitted because its latest dApp was only available on testnet and not yet accessible to all users.

Table~\ref{tab:protocols} reports the results of our selection process.
More specifically, it summarizes the information on the five most relevant derivative protocols ranked by TVL for the three protocol categories analyzed. 

\paragraph{Protocol Analysis}

For each of the protocols of Table~\ref{tab:protocols}, we conducted an in-depth manual analysis of their design and operational mechanisms, relying on the following sources:
\begin{itemize}
    \item official documentation such as whitepapers, developer documentation, and user guides;
    \item public interfaces such as dApps and websites;
    \item smart contract source code, when available via Etherscan, GitHub, or protocol explorers;
    \item online communities of the protocols, e.g., official Discord channels, and technical articles~\parencite{medium,dune,messari} for implementation-specific insights.
\end{itemize}
Table~\ref{tab:resr} in the appendix reports the entire list of sources we used for each analyzed protocol.

\paragraph{Framework Derivation}

Following an inductive approach, we derived the analytical framework from the protocol-level information collected in the previous phase. We first cleaned, filtered, compared, and harmonized the heterogeneous data gathered from available sources. We then abstracted from implementation-specific details to identify the recurring structural components that characterize derivative instruments across protocols.

This process led us to organize each instrument class around three interacting components: an instrument-level specification, a market state, and a protocol-level mechanism. The instrument-level component captures the contractual and economic attributes of a position. The market-state component captures the relevant price variables used for valuation, payoff determination, margin monitoring, and settlement. The protocol-level component captures the rules implemented by the protocol.

Based on this structure, we derive the main economic quantities associated with each derivative class. 

The resulting framework provides a protocol-agnostic abstraction for comparing heterogeneous derivatives protocols along common analytical dimensions, rather than a protocol-by-protocol account of implementation details.

\section{Analytical Representation of DeFi Derivative Instruments}
\label{sec:concept}

This section develops a unified analytical representation for the main derivative instruments in decentralized finance: perpetuals, options, and synthetic assets. Despite differences in payoff structure, lifecycle, and implementation, these instruments can be described through a common framework that separates instrument-level attributes, market-state variables, and protocol-level mechanisms.

For each instrument class, we adopt a tripartite representation, $\big(X, Y^{\mathrm{market}}, Z^{\mathrm{prot}}\big)$, where $X$ denotes the instrument specification, $Y^{\mathrm{market}}$ denotes the market state, and $Z^{\mathrm{prot}}$ denotes the protocol-level mechanism.

This representation provides a consistent basis for deriving the main economic quantities associated with each instrument class, such as unrealized profit and loss and collateralization ratios.

\subsection{Perpetuals}

Perpetual contracts are derivative instruments without a fixed expiration date, allowing traders to maintain positions indefinitely. They provide exposure to the price dynamics of an underlying asset without requiring contract rollover, as in traditional futures markets.
We represent a perpetual contract as three interacting components that together capture all the elements needed to describe its creation and evolution.

The instrument level specification of a position is represented by the tuple:
\[
X^{\mathrm{perp}} = \langle U, C, L, S, P_e \rangle,
\]
where $U$ denotes the underlying asset, $C$ denotes the collateral posted by the trader, $L$ denotes the leverage, $S \in \{+1,-1\}$ denotes the position direction, i.e., whether the position is long or short, and $P_e$ denotes the entry price.
Intuitively, the tuple captures the ingredients needed to open the position.

The protocol mechanism captures the dynamics affecting the position and can be formalized as the following tuple:
\[
Z^{\mathrm{prot}} = \langle \mathcal{F}, \tau^{\mathrm{tr}}, m_i, m_m \rangle,
\]
where $\mathcal{F}$ is the funding mechanism that defines how the funding rate $f$ is computed, $\tau^{\mathrm{tr}}$ denotes the transaction fee rate, and $m_i$ and $m_m$ denote the initial and the maintenance margin requirement, respectively.

The market state captures the relevant pricing variables and can be defined as:
\[
Y^{\mathrm{market}} = \langle P_{\mathrm{index}}, P_{\mathrm{mark}} \rangle,
\]
where $P_{\mathrm{index}}$ is an exogenous reference price constructed from external markets, and $P_{\mathrm{mark}}$ is the protocol-internal price used for valuation, margining, and liquidation. It is typically defined as a function of the index price and protocol-specific adjustments.

The triplet $\big( X^{\mathrm{perp}}, Y^{\mathrm{market}}, Z^{\mathrm{prot}}\big)$ provides a formal representation of the position specification, the relevant market variables, and the protocol rules governing position evolution.

\subsubsection{Position lifecycle}

Once opened, the position can be represented in time-indexed form. Some components of $X^{\mathrm{perp}}$, such as the underlying asset $U$, the position direction $S$, and the entry price $P_e$, remain fixed unless the position is modified. Other quantities, including equity, unrealized profit and loss, funding flows, notional exposure, and margin ratios, evolve as functions of the market state $Y^{\mathrm{market}}(t)$ and the protocol rules encoded in $Z^{\mathrm{prot}}$.

The initial notional exposure and the corresponding signed position size are determined by $C_0$ and $L_0$, the collateral and leverage chosen at position opening:
\[
N_0 = L_0 \cdot C_0, \qquad q = S \cdot \frac{N_0}{P_e},
\]
where $q$ represents the signed quantity of underlying asset exposure, e.g., the number of ETH units. This quantity remains constant over time unless the position is modified.
Unrealized PnL captures the mark-to-market variation at time $t$ of the position relative to the entry price:
\[
uPnL(t) = q \cdot \big(P_{\mathrm{mark}}(t) - P_e\big).
\]


The interaction between $P_{\mathrm{index}}(t)$ and $P_{\mathrm{mark}}(t)$ is central to the funding mechanism. In perpetual markets, the funding rate is typically linked to the relative deviation between the mark price, used for margining and liquidation, and the index price, which approximates the external spot price of the underlying asset. We denote by $\mathcal{F}$ a protocol-specific funding function:
\[
f(t) = \mathcal{F}\left(
\frac{P_{\mathrm{mark}}(t)-P_{\mathrm{index}}(t)}
{P_{\mathrm{index}}(t)}
\right).
\]

Funding is settled at discrete times $t_k$ according to the protocol mechanism $\mathcal{F}$, yielding a sequence of funding rates $f(t_k)$. Each settlement transfers value between long and short positions as a function of the position size, the mark price, and the funding rate. A positive funding rate implies that long positions pay funding, while short positions receive it. This formulation abstracts from protocol-specific adjustments, such as interest-rate components, clamps, borrowing fees, or price-impact terms, which can be incorporated into $\mathcal{F}$.

We denote by $\Phi^{\mathrm{fund}}(t)$ the cumulative funding component up to time $t$, obtained by aggregating all funding payments settled at times $t_k \le t$.

Account equity at time $t$ aggregates initial collateral, entry transaction costs, mark-to-market variation, and funding flows:
\[
E(t) = C_0 - I^{\mathrm{tr}} + uPnL(t) + \Phi^{\mathrm{fund}}(t),
\]
where $I^{\mathrm{tr}}$ denotes the entry transaction cost, defined as $I^{\mathrm{tr}} = \tau^{\mathrm{tr}} \cdot N_0$.



The current notional exposure is given by $N(t) = |q| \cdot P_{\mathrm{mark}}(t)$.
The margin ratio measures the solvency of the position relative to its current notional exposure:
\[
MR(t) = \frac{E(t)}{N(t)}.
\]

The position must satisfy the initial margin requirement at initiation, $MR(0) \ge m_i$, while at later times liquidation becomes admissible whenever the margin ratio falls below the maintenance threshold, $MR(t) \le m_m$.

\subsubsection{Protocol-level instantiations}

The protocol-level heterogeneity observed in practice can be interpreted as different instantiations of the components of the triplet $\big(X^{\mathrm{perp}}, Y^{\mathrm{market}}, Z^{\mathrm{prot}}\big)$, in particular with respect to supported underlying assets, collateral design, margin requirements, and fee mechanisms.


Table~\ref{tab:perps_protocols} reports information on the underlying assets and collateral available on the perpetual protocols investigated.

\begin{table}[htbp]
    \centering
    \scriptsize
    \caption{Overview of selected perpetual protocols (snapshot: \textit{April 2025}). 
    Supported Assets are grouped by categories; each cell encodes (underlying, collateral) support from left to right:
    $\bullet\!\circ$ = underlying only, $\circ\!\bullet$ = collateral only, $\bullet\!\bullet$ = both, $\circ\!\circ$ = not supported.}
    \setlength{\tabcolsep}{3pt} 
    \begin{tabular*}{0.98\linewidth}{@{\extracolsep{\fill}}lcccccccccc@{}}
        \toprule
        \textbf{Protocol} & \multicolumn{8}{c}{\textbf{Supported Assets}} & \textbf{\#U} & \textbf{\#C} \\ 
        \cmidrule(lr){2-9}
         & L1 & L2 & DeFi & Meme & Gaming & Forex & RWA & Stable  &  &  \\ 
        \midrule
        Jupiter Exchange 
            & \assetUC  & \assetNone & \assetNone & \assetNone & \assetNone & \assetNone & \assetNone & \assetC
            & 3  
            & 5 \\
        Drift Trade 
            & \assetUC  & \assetU    & \assetU    & \assetU    & \assetNone & \assetNone & \assetNone & \assetC
            & 49  
            & 2  \\ 
        GMX (Arbitrum) 
            & \assetUC  & \assetUC   & \assetUC   & \assetUC   & \assetUC   & \assetNone & \assetNone & \assetC
            & 49 
            & 49  \\
        dYdX 
            & \assetU   & \assetU    & \assetU    & \assetU    & \assetU    & \assetU    & \assetU    & \assetUC
            & 671 
            & 1  \\
        Hyperliquid 
            & \assetUC  & \assetUC   & \assetUC   & \assetUC   & \assetUC   & \assetNone & \assetNone & \assetC
            & 45 
            & 45  \\
        \bottomrule
    \end{tabular*}
    \label{tab:perps_protocols}
\end{table}

In the table, the term ``L1'' refers to native blockchain cryptocurrencies such as SOL (the native cryptocurrency of the Solana blockchain), BTC (the Bitcoin cryptocurrency), and ETH (the native cryptocurrency of Ethereum). The term ``L2'' refers to tokens associated with Layer~2 protocols~\parencite{GangwalGT23}, i.e., blockchain protocols built on top of existing blockchain platforms (e.g., Ethereum as a base layer) to address scalability or transaction-cost limitations. The label ``DeFi'' includes governance and utility tokens of decentralized finance protocols, such as AAVE~(\citeyear{aave}), LDO~(\citeyear{lido}), and UNI~(\citeyear{uniswap}). 
The category ``Meme'' refers to tokens inspired by Internet memes or cultural references (e.g., Dogecoin or Pepe), while ``Gaming'' collects tokens used within blockchain-based games and metaverse environments. The category ``Forex'' comprises instruments whose payoff is linked to traditional foreign-exchange rates (e.g., TRY/USD or EUR/USD), and ``RWA'' (Real-World Assets) denotes tokens representing on-chain claims on off-chain assets \parencite{vella2026taxonomyrealworldassettokenization}. Finally, ``Stable'' refers to cryptocurrencies designed to maintain a relatively stable value relative to an external reference, typically a fiat currency.

Table~\ref{tab:perps_protocols} highlights the heterogeneous design choices adopted by decentralized perpetual protocols. 
Jupiter Exchange lists a limited set of Layer~1 assets as underlyings, while allowing stablecoins as collateral only. Drift Trade and GMX exhibit a broader coverage of underlying assets but differ significantly in collateral design: Drift restricts collateral to a small set of assets, whereas GMX allows most listed assets to be used as margin. Hyperliquid follows a similar approach to GMX. 
dYdX supports by far the broadest universe of underlying assets, spanning all categories, including less standard instruments such as forex pairs and tokenized real-world assets. Despite this extensive coverage, collateral is restricted to a single stablecoin (USDC), reflecting a conservative collateral policy.
These differences suggest that protocols tend to restrict collateral to highly liquid, widely recognized assets, while allowing greater flexibility on the underlying side.


\begin{table}[htbp]
	\caption{Maintenance-margin rules and maximum leverage offered by leading perpetual protocols}
	\label{tab:perpmargins}
	\centering
	\begin{tabular}{l p{6cm} p{4cm}}
		\toprule
		\textbf{Platform} & \textbf{Maintenance-margin policy} & \textbf{Maximum leverage} \\
		\midrule
        
		Jupiter Exchange &
		Maintenance margin is computed from the net exposure after fees:  
		\(\displaystyle \text{Mm}= \text{price} \pm 
		\frac{\lvert \text{C} - \text{close fee} - \text{borrow fee} - \text{size}/\text{L} \rvert \times \text{price}}
		{\text{size}}\),
		with “\(+\)” for shorts and “\(-\)” for longs. &
		100\,$\times$ on Solana; 150\,$\times$ on Ethereum and Wrapped Bitcoin. \\[4pt]
		
		Drift &
		Asset-specific schedule: maintenance margin ranges from 3 \% (most liquid) to 16.67 \% (least liquid). &
		101\,$\times$ on SOL, ETH, BTC; 3–10\,$\times$ on other tokens. \\[4pt]
		
		GMX &
		Not disclosed in the public documentation. &
		100\,$\times$ on major tokens; 50\,$\times$ on others. \\[4pt]
		
		dYdX &
		Asset-dependent range between 0.05 \% and 10 \%. &
		50\,$\times$ on BTC and ETH; 20\,$\times$ on SOL; 5–10\,$\times$ on other tokens. \\[4pt]
		
		Hyperliquid &
		Maintenance margin equals one-half of the initial margin at max leverage. With max leverage from 3 × to 40 ×, this yields 16.7 \% to 1.25 \%. &
		3–40\,$\times$, depending on the asset. \\
		\bottomrule
	\end{tabular}
\end{table}

Protocols also differ in their specification of margin constraints, which directly instantiate the parameters $m_i$ and $m_m$ in the protocol-level component $Z^{\mathrm{prot}}$.
Table~\ref{tab:perpmargins} summarizes maintenance-margin rules and maximum leverage across protocols.
Interestingly, maximum leverage varies significantly across platforms, ranging from 3$\times$ to above 100$\times$, implying substantial differences in attainable exposure and risk profiles. Margin requirements are often asset-dependent, reflecting differences in liquidity and volatility across traded instruments. Some protocols adopt explicit formulas based on position size and fees (e.g., Jupiter), while others define tiered or asset-specific schedules (e.g., Drift, dYdX).


\begin{table}[htbp]
	\caption{Fee structure of leading perpetual protocols}\label{tab:perpfees}
	\centering
	\scriptsize
	\begin{tabular}{l p{5.5cm} p{5.5cm}}
		\toprule
		\textbf{Platform} & \textbf{Ongoing hourly fees} & \textbf{One-off fees (open/close)} \\
		\midrule
		Jupiter Exchange &
		Borrow fee only: hourly borrow fee = utilization ratio × hourly borrow rate × position size.  
		No funding fee. &
		Flat 0.06 \% of notional each time a position is opened or closed; extra \emph{price-impact fee} when the trade moves the pool price. \\[4pt]
		
		Drift &
		Funding fee: every hour the trader pays/receives one-twenty-fourth of the percentage gap between the 1-h EWMA mid-price and the oracle reference price. &
		Market-specific base fee 0.03–0.10 \%; additional “insurance / borrow” surcharge 0.75–5.0 \%. \\[4pt]
		
		GMX &
		Single hourly charge that combines funding and borrow components (formula in protocol docs). &
		Base fee 0.04–0.06 \% of notional, discounted by volume tiers; \emph{price-impact fee} for large orders. \\[4pt]
		
		dYdX &
		Funding premium: \((\max(0,\text{Bid}-\text{Index})-\max(0,\text{Index}-\text{Ask}))/\text{Index}\). &
		Base fee 0.025–0.05 \% with volume tiers; ongoing trader-rewards programme can offset part of the cost. \\[4pt]
		
		Hyperliquid &
		Hourly funding payment: funding rate = average premium index + \(\text{clamp}(\text{interest rate}-\text{premium},-0.0005,0.0005)\). &
		Maker–taker tier schedule: 0–0.045 \% of notional. \\
		\bottomrule
	\end{tabular}
\end{table}

The fee structures reported in Table~\ref{tab:perpfees} correspond to different specifications of the funding mechanism $\mathcal{F}$, and the transaction cost parameter $\tau^{\mathrm{tr}}$ in $Z^{\mathrm{prot}}$.
While most protocols implement periodic funding payments to align the perpetual price with the underlying index, the specific design of these mechanisms varies substantially. Some platforms rely on explicit funding rates computed from price deviations (e.g., Drift, dYdX, Hyperliquid), whereas others combine funding and borrowing components into a unified fee (e.g., GMX), or replace funding altogether with borrowing fees (e.g., Jupiter).
One-off transaction costs, typically proportional to traded notional, are consistently observed across protocols, although their magnitude and structure depend on volume tiers, market conditions, and additional factors such as price impact.
While the analytical representation abstracts from implementation-specific details, these heterogeneous mechanisms can be interpreted as different realizations of a generalized funding function and transaction cost structure within the same formal framework.

\subsection{Options}

Options are derivative instruments that grant the holder the right, but not the obligation, to buy or sell an underlying asset at a predetermined price. In decentralized finance, they appear in both expiring and non-expiring forms. 
While expiring options concentrate their economic logic at maturity, everlasting options introduce mechanisms that make their behavior closer to that of perpetual contracts. 

We represent option-based instruments through the same triplet $\big( X, Y^{\mathrm{market}}, Z^{\mathrm{prot}}\big)$.

\subsubsection{Expiring options}

At the instrument level, an expiring option is represented as the following tuple:
\[
X^{\mathrm{opt}}=\langle U, C, L, S, P_{\text{s}}, T_{\text{expiry}} \rangle,
\]
where $U$ is the underlying asset, $C$ is the collateral, $L$ is the leverage factor when margining is supported, $S$ denotes the option type, $P_{\text{s}}$ is the strike price, and $T_{\text{expiry}}$ is the expiration date.
In this context, the strategy variable $S$ distinguishes between the two standard payoff structures:
\begin{itemize}
    \item \emph{Call option}: the right to buy the underlying asset at the strike price;
    \item \emph{Put option}: the right to sell the underlying asset at the strike price.
\end{itemize}
Again, the tuple represents the ingredients to open a position.

The market state is captured by the underlying price, resulting in a 1-element tuple:
\[
Y^{\mathrm{market}}=\langle P_U\rangle.
\]

The protocol component $Z^{\mathrm{prot}}$ is typically minimal for fully collateralized options, but may include margin requirements or liquidation rules in margin-based implementations.

Also in this case, once the position is opened, the tuple becomes a function of the time $X^{\mathrm{opt}}(t)$.
The economic logic of expiring options is concentrated at maturity. Let $P_U(T_{\mathrm{expiry}})$ denote the underlying price at expiration. The payoff of one option contract is:
\[
\Pi_T=
\begin{cases}
\max(P_U(T_{\mathrm{expiry}})-P_{\mathrm{s}},0), & \text{if } S=\mathrm{call},\\[4pt]
\max(P_{\mathrm{s}}-P_U(T_{\mathrm{expiry}}),0), & \text{if } S=\mathrm{put}.
\end{cases}
\]

This representation highlights that expiring options do not rely on intertemporal funding or margin dynamics in their simplest form, as their valuation is primarily determined by the strike price and the contract horizon.

\subsubsection{Everlasting options}

Everlasting options are derivative instruments that provide option-like exposure without a fixed expiration date. Their structure extends that of expiring options by replacing the maturity constraint with a funding-based mechanism that regulates the persistence of the position over time.

At the instrument level, we represent an everlasting option as
\[
X^{\mathrm{eopt}}=\langle U, C, L, S, P_{\text{s}} \rangle,
\]
where the components retain the same interpretation as in expiring options, except for the absence of a fixed expiration date.
The market-state component is analogous to that of expiring options, while the protocol-level component introduces a funding mechanism and margin requirements:
\[
Z^{\mathrm{prot}}=\langle \mathcal{F}, m_i, m_m \rangle,
\]
where $\mathcal{F}$ denotes the discrete funding mechanism and $m_i, m_m$ represent initial and maintenance margin constraints.

Given this structure, the economic behavior of everlasting options combines elements of both expiring options and perpetual contracts. As in expiring options, the payoff structure depends on the strike price and option type. However, the absence of maturity implies that positions are subject to funding payments and margin constraints, analogous to perpetual derivatives. For this reason, everlasting options can be interpreted as an intermediate class of instruments, bridging the static payoff structure of expiring options and the continuous-time exposure and funding dynamics of perpetual contracts.\footnote{A full treatment of option pricing lies beyond the scope of this study. For protocol-specific pricing approaches for perpetual or everlasting options, see e.g.~\parencite{deri:protocol:pricing}.}

\subsubsection{Protocol-level instantiations}

Table~\ref{tab:opt_protocols} reports the underlying assets and collateral supported by the options protocols we consider in this study.
Compared to perpetuals, options are limited to a smaller number of cryptoassets, primarily stablecoins such as USDC and USDT, or major native cryptoassets like ETH and BTC. Additionally, protocols offering expiring options focus mainly on these two underlyings, ETH and BTC. In contrast, perpetual contracts include a wider variety of underlying assets and different types of collateral.

\begin{table}[htbp]
\centering
\footnotesize
\caption{Overview of selected option protocols: underlying assets, collateral assets, and option type.}
\label{tab:opt_protocols}
{
  \setlength{\tabcolsep}{3pt} 
  
  \begin{tabular*}{0.98\linewidth}{@{\extracolsep{\fill}}l
                                   p{0.23\linewidth}
                                   p{0.45\linewidth}
                                   c@{}}
  \toprule
  \textbf{Protocol} & \textbf{Underlying} & \textbf{Collateral} & \textbf{Option Type} \\
  \midrule
  Derive    & ETH, BTC                       & USDC                                                 & Expiring    \\
  Hegic     & ETH, BTC                       & USDC.e                                               & Expiring    \\
  SOFA.org  & ETH, BTC                       & USDT, crvUSD, stETH, RCH                             & Expiring    \\[0.5ex]
  \midrule
  Deri V4   & SOL, ETH, BTC, BNB, TON, SUI   & USDC, ETH, USDT, WBTC, DAI, ARB, LINK, DERI, LUSD, wstETH & Everlasting \\ 
  GammaSwap & weETH, WETH, WBTC, PENDLE, PEPE, ARB, GS
            & WETH, USDC, USDC.e, USD0++, WBTC                & Everlasting \\
  \bottomrule
  \end{tabular*}
}
\end{table}

\subsection{Synthetics}

Synthetic assets are digital financial instruments that replicate the value or price exposure of an underlying asset without requiring direct ownership. In decentralized finance, this exposure is typically implemented through over-collateralized minting mechanisms, whereby users lock collateral to issue synthetic tokens tracking a reference asset.
We represent synthetic instruments through the triplet  $\big(X^{\mathrm{synth}}, Y^{\mathrm{market}}, Z^{\mathrm{prot}}\big)$.

At the instrument level, a synthetic asset is defined as the tuple
\[
X^{\mathrm{synth}}=\langle U, C \rangle,
\]
where $U$ is the underlying asset to be tracked and $C$ is the collateral posted to support minting. The market component captures the evolution of the relevant price processes:
\[
Y^{\mathrm{market}}=\langle P_U, P_C \rangle,
\]
where $P_U$ denotes the reference price of the underlying and $P_C$ the price of the collateral.

The protocol component encodes the collateralization and fee structure:
\[
Z^{\mathrm{prot}}=\langle \gamma, \phi^{\mathrm{stab}} \rangle,
\]
where $\gamma$ is the minimum collateralization threshold and $\phi^{\mathrm{stab}}$ represents stability or borrowing fees applied to open positions.

\subsubsection{Position lifecycle}

Again, once opened, we represent the position as a function over time that depends also on the market conditions $Y^{\mathrm{market}}$ and the protocol mechanism $Z^{\mathrm{prot}}$. We denote this function $X^{\mathrm{synth}}(t)$ that corresponds to a time-indexed instantiation of the tuple above. 

Given this structure, the core economic variables of a synthetic position can be derived directly. Let $M(t)$ denote the number of synthetic units minted at time $t$. The corresponding debt value is:
\[
D(t) = M(t)\, P_U(t),
\]
which reflects the obligation of the user denominated in the reference value of the underlying. The value of posted collateral is:
\[
V^{C}(t)=C(t)\, P_C(t),
\]
where $C(t)$ denotes the collateral balance at time $t$.

The collateralization ratio, which is the key solvency metric in synthetic protocols, is then defined as:
\[
CR(t)=\frac{V^{C}(t)}{D(t)}=
\frac{C(t)P_C(t)}{M(t)P_U(t)}.
\]
Protocol safety requires this ratio to remain above the minimum threshold $\gamma$. Liquidation becomes admissible whenever
$CR(t)\le \gamma$.

In contrast to perpetual contracts, synthetic protocols do not rely on directional leverage or funding transfers between long and short positions. Instead, their internal consistency is ensured through over-collateralization, debt accounting, and liquidation mechanisms. 

While the analytical representation abstracts from implementation-specific details, different protocols instantiate these components through heterogeneous choices of collateral types, supported underlyings, and fee structures.

\subsubsection{Protocol-level instantiations}

Table~\ref{tab:synth_protocols} reports the underlying assets and collateral types supported by the synthetic protocols analyzed in this study, providing a concrete illustration of how the components $(U, C)$ and the associated collateral policies are instantiated in practice.

\begin{table}[htbp]
\centering
\scriptsize
\caption{Overview of selected synthetic protocols: underlying assets and collateral assets.}
\label{tab:synth_protocols}
{
  \setlength{\tabcolsep}{3pt} 

  \begin{tabular*}{0.98\linewidth}{@{\extracolsep{\fill}}l
                                   p{0.30\linewidth}
                                   p{0.54\linewidth}@{}}
  \toprule
  \textbf{Protocol} & \textbf{Underlying} & \textbf{Collateral} \\
  \midrule
  Synthetix       & sUSD, sETH, sBTC, sLINK 
                  & Layer 1, Layer 2, DeFi \\

  Alchemix        & alUSD, alETH 
                  & ETH, WETH, DAI, USDC, USDT, FRAX \\

  Youves          & uUSD, uBTC, uXTZ, uXAU, uDEFI 
                  & TEZ, tzBTC, SIRS \\

  Metronome Synth & msUSD, msETH, msBTC, msOP 
                  & ETH, WBTC, DAI, USDC, FRAX, sfrxETH, vaETH, vaUSDC, vaFRAX, vacbETH, varETH, vastETH, OP, vaOP \\[3pt]

  Taiga           & tDOT 
                  & DOT, LDOT \\
  \bottomrule
  \end{tabular*}
}
\end{table}

Synthetix V3 supports assets such as sUSD, sETH, sBTC, and sLINK. Legacy assets, including sAMZN, sTSLA, and sAAPL, are no longer actively used and are therefore excluded from our table. A more comprehensive list is available on Etherscan.\footnote{\url{https://etherscan.io/tokens/label/synthetix?subcatid=3-0&size=7&start=0&col=3&order=desc}}

\section{DeFi Derivative Protocols: Agents, Components, and Operational Dynamics} \label{sec:protocols}

So far, we have introduced the main derivative instrument classes and their analytical representation in terms of the triplet 
$\big(X, Y^{\mathrm{market}}, Z^{\mathrm{prot}}\big)$, 
which abstracts instrument-level features, market-state variables, and protocol-level mechanisms.

This section translates the analytical representation into its operational counterparts: agents instantiate instrument-level choices, oracles provide market-state variables, and smart-contract modules enforce protocol-level rules.

Our conceptualization highlights that perpetual and option protocols involve similar agents, components, and lifecycle dynamics; accordingly, we analyze them jointly. Synthetic protocols are then examined separately, since their operational logic differs in that exposure is generated through collateralized minting and debt creation rather than through margined trading positions.

\begin{table}[htbp]
  \caption{Summary of the main economic agents and protocol components involved in decentralized derivatives protocols, and their relation to the analytical framework.}
  \label{tab:glossary2}
  \renewcommand{\arraystretch}{1.35}
  
  \begin{tabular}{p{0.28\textwidth}p{0.67\textwidth}}
    \toprule
    \textbf{Category} & \textbf{Definition} \\
    \midrule
    
    \multicolumn{2}{l}{\textbf{Economic Agents}} \\
    \midrule
    
    Trader & A user who creates, modifies, and closes derivative positions by selecting contractual parameters (e.g., underlying, leverage, strike, collateral), thereby instantiating the instrument-level component $X$. \\
    
    Liquidity Provider (LP) & A user who supplies capital to the protocol, typically through liquidity pools, supporting leveraged trading or pool-based counterparty exposure, and earning a share of fees or incentives. \\
    
    Liquidator / Keeper & An external agent who monitors protocol-defined solvency conditions and triggers liquidation procedures when margin or collateral constraints are violated. \\
    
    \midrule
    \multicolumn{2}{l}{\textbf{Protocol Components}} \\
    \midrule
    
    Liquidity Pool & A smart-contract-based reserve of assets used to support trading, leverage, settlement, or counterparty exposure, depending on the protocol design. \\
    
    Oracle & A component that provides the reference data required to determine the market state $Y^{\mathrm{market}}$, including prices used for valuation, risk monitoring, and liquidation. \\
    
    Position-Management / \\ Execution Module & A protocol component that governs order submission, matching, execution, and the registration of position updates in protocol state, implementing the transition from user intent to realized exposure. \\

    Clearing and \\ Risk-Management Module & A protocol-level component, or set of smart contracts, that operationalizes $Z^{\mathrm{prot}}$ by computing profit and loss, enforcing margin or collateral constraints, tracking account health, and triggering liquidation procedures. \\
    
    \bottomrule
  \end{tabular}
\end{table}

\subsection{Agents and components in perpetual and option protocols} 
\label{perpopt}

The agents and components involved in perpetual and option protocols can be interpreted as the operational counterparts of the analytical framework introduced in Section~\ref{sec:concept}. Collectively, they determine how instrument specifications are instantiated, how market-state variables are observed, and how protocol rules are enforced throughout the lifecycle of a position.
Three main economic agents participate in protocol activity: traders, liquidity providers, and liquidators or keepers.\footnote{
Governance users, i.e. holders of governance tokens that grant voting rights and decision-making power, also play a crucial role in protocol activity. However, a detailed discussion of their role goes beyond the scope of this study. For further details, see e.g.~\parencite{kitzler2024governance,han2025review}.
}
Table~\ref{tab:glossary2} summarizes these agents and protocol components.

\textit{Traders} are the primary agents through which the instrument-level component $X$ is instantiated. By selecting contractual parameters such as the underlying asset, position type, leverage, strike, or collateral structure, they determine the economic configuration of the position.

\textit{Liquidity providers} (LPs) are users who supply assets to a \textit{liquidity pool}, i.e. one or more smart contracts used to store reserves of crypto-assets. These reserves can serve several purposes, including supporting leverage-based trading activity and acting as a counterparty in pool-based market designs. In return for providing liquidity, LPs may receive a liquidity-pool token representing their share of the pool’s assets and entitling them to a proportional share of fees and other protocol incentives. The specific reward structure varies across protocols.

\textit{Liquidators} or keepers are external agents who monitor the health of open positions and start liquidation procedures when positions are undercollateralized by activating predefined smart-contract functions once collateral or margin conditions fall below the required threshold.

Among the main protocol components, \textit{oracles} provide the exogenous inputs required to determine the market state $Y^{\mathrm{market}}$, supplying the price data used for valuation, risk monitoring, and liquidation.\footnote{Oracles may also derive data from other smart contracts or decentralized exchanges. In that case, they are often referred to as on-chain oracles. Off-chain oracles instead pull data from external sources, such as traditional financial markets or web APIs.}
Oracles may differ significantly in their trust assumptions~\parencite{al-breiki:trustworthy:2020,}: centralized solutions depend on a single data provider and offer efficiency at the cost of greater concentration risk, whereas decentralized oracles, such as Chainlink~(\citeyear{chainlink:oracle}) and Band Protocol~(\citeyear{band:oracle}), aggregate information from multiple independent nodes, thereby improving robustness and trustworthiness.

The \textit{position-management / execution module} governs how orders are submitted, matched, and recorded in protocol state. It implements the transition from user intent to realized exposure. Across the protocols considered, this module is implemented through different execution and counterparty structures. In the \emph{order-book} model, trades are matched on a continuous limit order book, either on-chain or off-chain, and price discovery is endogenous to the order flow. In the \emph{pool-counterparty} model, by contrast, users trade against a liquidity pool rather than against another trader, and execution prices are typically determined using external reference prices together with protocol-specific price-impact and fee rules. Some protocols also adopt hybrid structures that combine order-book execution with additional protocol-level mechanisms for margining, settlement, or liquidity support.

The \textit{clearing and risk-management module} operationalizes $Z^{\mathrm{prot}}$ by enforcing margin requirements, computing profit and loss, tracking account health, and triggering liquidation.

\begin{table}[htbp]
\centering
\footnotesize
\label{tab:match_engine}
\caption{Execution model, counterparty structure, and oracle design across selected derivatives protocols}
\begin{tabular}{p{1.3cm} p{2.5cm} p{4.3cm} p{4.3cm}}
\toprule
\textbf{Protocol} & \textbf{Execution model} & \textbf{Liquidity pool} & \textbf{Oracle} \\
\midrule
Jupiter Exchange & pool--counterparty & Single index pool (SOL, ETH, WBTC, USDC, USDT); pool PnL and a share of fees accrue to LP & Edge (primary); Chainlink and Pyth for verification/backup \\
Drift & Hybrid & No single counterparty pool. & Pyth \\
GMX & pool--counterparty & Per-market GM pools; optional GLV vaults allocate across markets; long/short tokens back positions & Chainlink  \\
dYdX v4 & Order book & No counterparty pool & Chain-native validator-aggregated oracle \\
Hyperliquid & Order book & No counterparty pool & Weighted median of major venues (Binance, OKX, Bybit, Kraken, KuCoin, Gate.io, MEXC, Hyperliquid) \\
\midrule
Derive & Order book & No counterparty pool & Black Scholes oracle \\
Hegic & pool--counterparty & Single-asset pools (ETH, WBTC) & Chainlink \\
SOFA.org & Market-makers  & No counterparty pool & Chainlink  \\
Deri V4 & pool--counterparty & Unified cross-chain pool & Oraclum and Pyth \\
GammaSwap & pool--counterparty & Borrows LP tokens from external pools & Oracleless \\
\bottomrule
\end{tabular}
\end{table}

\subsection{Dynamics of perpetual and option protocols} 
\label{sec:dyn}

\begin{figure}[!htbp]
\centering
\begin{adjustbox}{max totalsize={\textwidth}{0.88\textheight},center}
\begin{tikzpicture}[
font=\small,
participant/.style={
draw,
rounded corners,
fill=gray!10,
minimum width=2.7cm,
minimum height=0.75cm,
align=center
},
lifeline/.style={draw=gray!55, dashed},
msg/.style={-stealth, thick},
ret/.style={-stealth, thick, dashed},
phase/.style={
draw=gray!40,
fill=gray!12,
rounded corners,
minimum width=16.2cm,
minimum height=0.55cm,
font=\bfseries
},
fragment/.style={
draw=gray!45,
rounded corners=2pt,
fill=gray!3,
fill opacity=0.25,
text opacity=1,
line width=0.35pt
},
fragmentlabel/.style={
draw=gray!45,
fill=white,
fill opacity=0.85,
text opacity=1,
rounded corners=1pt,
font=\scriptsize\bfseries,
inner xsep=4pt,
inner ysep=2pt
},
fragmentcond/.style={
font=\scriptsize\itshape,
anchor=west,
text opacity=1
}
]

\node[participant] (TR)   at (0,0)    {Trader};
\node[participant] (OB)   at (3.2,0)  {Position Management};
\node[participant] (ORA)  at (6.4,0)  {Oracle};
\node[participant] (COL)  at (9.6,0)  {Collateral Vault};
\node[participant] (CH)   at (12.8,0) {Clearing House};
\node[participant] (POOL) at (16.0,0) {Liquidity Pool};

\foreach \p in {TR,OB,ORA,COL,CH,POOL}
\draw[lifeline] (\p.south) -- ++(0,-21.8);

\node[phase] at (8,-1.0) {Opening the position};

\draw[ret]
(TR|-0,-1.8) --
node[above, sloped] {\scriptsize 1. Submit order}
(OB|-0,-1.8);

\draw[ret]
(OB|-0,-2.5) --
node[above, sloped] {\scriptsize 2. Request reference price}
(ORA|-0,-2.5);

\draw[msg]
(TR|-0,-3.2) --
node[above, sloped] {\scriptsize 3. Provide collateral}
(COL|-0,-3.2);

\draw[msg]
(OB|-0,-3.9) --
node[above, sloped] {\scriptsize 4. Validate risk constraints}
(CH|-0,-3.9);

\draw[ret]
(CH|-0,-4.6) --
node[above, sloped] {\scriptsize 5. Read reference price}
(ORA|-0,-4.6);

\draw[ret]
(CH|-0,-5.3) --
node[above, sloped] {\scriptsize 6. Check posted collateral}
(COL|-0,-5.3);

\draw[fragment]
(-0.45,-6.05) rectangle (16.45,-11.45);

\node[fragmentlabel] at (0.0,-6.05) {alt};
\node[fragmentcond] at (0.45,-6.35) {[risk constraints satisfied]};

\draw[ret]
(CH|-0,-7.2) --
node[above, sloped] {\scriptsize 7. Approve position opening}
(OB|-0,-7.2);

\draw[msg]
(POOL|-0,-7.9) --
node[above, sloped] {\scriptsize 8. Provide liquidity for leverage}
(OB|-0,-7.9);

\draw[msg]
(TR|-0,-8.6) --
node[above, sloped] {\scriptsize 9. Settle trading fee}
(POOL|-0,-8.6);

\draw[msg]
(OB|-0,-9.3) --
node[above, sloped] {\scriptsize 10. Create position}
(TR|-0,-9.3);

\draw[gray!40, dashed, line width=0.3pt]
(-0.45,-9.85) -- (16.45,-9.85);

\node[fragmentcond] at (0.45,-10.15) {[otherwise]};

\draw[msg]
(CH|-0,-11.0) --
node[above, sloped] {\scriptsize 7'. Reject order / request more collateral}
(TR|-0,-11.0);

\node[phase] at (8,-12.3) {Position monitoring};

\draw[msg]
(TR|-0,-13.1) --
node[above, sloped] {\scriptsize 11. Modify position}
(OB|-0,-13.1);

\draw[msg]
(OB|-0,-13.8) --
node[above, sloped] {\scriptsize 12. Revalidate risk constraints}
(CH|-0,-13.8);

\draw[fragment]
(-0.45,-14.45) rectangle (16.45,-18.95);

\node[fragmentlabel] at (0.0,-14.45) {alt};
\node[fragmentcond] at (0.45,-14.75) {[compliant]};

\draw[msg]
(CH|-0,-15.8) --
node[above, sloped] {\scriptsize 13. Keep position open}
(TR|-0,-15.8);

\draw[gray!40, dashed, line width=0.3pt]
(-0.45,-16.45) -- (16.45,-16.45);

\node[fragmentcond] at (0.45,-16.75) {[under-collateralized]};

\draw[ret]
(CH|-0,-17.8) --
node[above, sloped] {\scriptsize 13'. Execute liquidation}
(COL|-0,-17.8);

\draw[msg]
(CH|-0,-18.6) --
node[above, sloped] {\scriptsize 14. Return residual collateral, if any}
(TR|-0,-18.6);

\node[phase] at (8,-19.9) {Closing the position};

\draw[ret]
(TR|-0,-20.7) --
node[above, sloped] {\scriptsize 15. Submit close request}
(OB|-0,-20.7);

\draw[msg]
(OB|-0,-21.5) --
node[above, sloped] {\scriptsize 16. Realize PnL}
(TR|-0,-21.5);


\end{tikzpicture}
\end{adjustbox}
\caption{Sequence diagram of the main interactions involved in opening, monitoring, liquidating, and closing a leveraged perpetual position. Dashed arrows indicate informational interactions without direct value transfer, whereas solid arrows denote value transfers or state-changing actions.}
\label{fig:general_schema}
\end{figure}

We now provide an overview of the operational dynamics of perpetual and option derivatives protocols.
A typical derivatives protocol exhibits two main operational flows. The first is the \emph{liquidity provision flow} that captures the process through which liquidity providers supply and withdraw capital. The second is the \emph{trading flow}, shown in Figure~\ref{fig:general_schema}, which encompasses the sequence of actions a trader follows to open, manage, and close positions. 
The following subsections describe the exchange models and these operational flows in detail.

\subsubsection{Liquidity provision flow}

The liquidity flow specifies how a protocol governs the supply, utilization, and withdrawal of assets, and clarifies the structure of liquidity provisioning as well as the role of liquidity providers (LPs) in market operations.

LPs contribute assets to a liquidity pool and receive LP tokens that represent a claim on the pool's net asset value. Two accounting designs are common. 
For reward-bearing tokens, the LP token's unit value appreciates as fees and rewards accumulate; instead, in the case of rebase tokens, the protocol increases the token supply to reflect increased rewards.
For example, in Jupiter~(\citeyear{jlp}), appreciation is driven by protocol-generated fees that accumulate to the pool, allowing the LP token's value to increase over time.

Finally, liquidity withdrawal proceeds via a burn-and-redeem mechanism: LP tokens are burnt and the protocol returns the corresponding share of the pool. Depending on the implementation, withdrawals may be subject to fees or limits to preserve pool stability and reduce liquidity shocks.

\subsubsection{Trading flow}
This second mechanism describes how a position is opened, monitored, and closed within a protocol. We decompose this flow into three main phases: $(i)$ \emph{opening the position}, $(ii)$ \emph{monitoring and managing the position}, and $(iii)$ \emph{closing the position}.  
At position opening, the trader instantiates $X$, the oracle supplies the variables entering $Y^{\mathrm{market}}(t)$, and the protocol validates the constraints encoded in $Z^{\mathrm{prot}}$ before execution.
Then, positions are continuously updated as market prices evolve, affecting $Y^{\mathrm{market}}(t)$ and consequently equity, margin ratios, and funding flows. Finally, closing a position finalizes the realized PnL and settles all associated protocol-level cash flows.
Below, we detail these phases and analyze how trading operations interact with the protocol’s infrastructure.

\paragraph{Opening the position}

A trader initiates a position by submitting an order through the protocol’s position-management module (Action~1, Fig.~\ref{fig:general_schema}). The protocol then requests the current reference price from the oracle (Action~2), while the trader provides the required collateral (Action~3). Before the position can be created, the protocol validates the relevant risk constraints (Action~4), using current price information and the posted collateral as inputs (Actions~5--6). This validation determines whether the proposed position satisfies the protocol’s requirements in terms of collateral adequacy, leverage, and margin conditions. If these constraints are satisfied, the protocol approves the opening of the position (Action~7). When leverage is involved, additional liquidity is then sourced from the liquidity pool (Action~8), and the trader settles the applicable trading fee (Action~9). The position is subsequently created and recorded in protocol state (Action~10). If the required risk constraints are not satisfied, the protocol rejects the order or requests additional collateral instead (Action~7', alternative branch).

At entry, the position’s economic exposure depends on posted collateral, leverage, and contract specifications. The notional exposure $N(t)$ serves as the reference base for several computations, including trading fees, borrowing costs, and funding transfers, and it updates with size adjustments over the life of the position. In pool-based designs, leverage is supported by protocol liquidity, while borrowing and trading fees are typically charged upon opening and during the life of the position.

\paragraph{Monitoring and managing the position}

Once opened, a position enters a monitoring phase in which the trader may submit a modification request, for instance to change the position’s exposure or collateralization level, when supported by the protocol (Action~11). This update triggers a renewed validation of the protocol’s risk constraints (Action~12). If the position remains compliant, it is kept open (Action~13, compliant branch). If instead the position becomes under-collateralized, liquidation may be executed (Action~13, liquidation branch), and any residual collateral is returned to the trader when available (Action~14).

More generally, the protocol continuously monitors account health as market prices evolve. Positions are marked to market using updated reference prices, and account equity changes with unrealized profit and loss, accrued funding transfers, borrowing costs, execution fees, and collateral movements. In this way, the protocol repeatedly checks whether the position remains above the relevant maintenance threshold. We abstract here from implementation-specific details such as partial liquidation mechanisms or protocol-specific auction procedures.

\paragraph{Closing the position}
A position can be closed by the trader through the protocol’s position-management module (Action~15). At that point, the protocol computes and realizes the final profit and loss (Action~16), which is then reflected in the trader’s account balance. More complex closing conditions, such as stop-loss or take-profit rules, can be understood as protocol-specific variations of this general closing logic.

\subsection{Synthetic protocols} \label{protocol_synths}
Synthetic protocols generate exposure by minting a synthetic asset against posted collateral, thereby creating a collateralized debt position that must remain sufficiently over-collateralized over time. While they share some core components with perpetual and option protocols, such as oracles, collateral vaults, and clearing-house mechanisms, their operational logic differs in that exposure is created through minting and debt issuance rather than through margined trading positions. Figure~\ref{fig:synthetic-position-sequence} summarizes this lifecycle by distinguishing three main phases: minting, position monitoring, and redemption.

\begin{figure}[!htbp]
\centering
\begin{adjustbox}{max totalsize={\textwidth}{0.88\textheight},center}
\begin{tikzpicture}[
    font=\small,
    participant/.style={
        draw,
        rounded corners,
        fill=gray!10,
        minimum width=2.8cm,
        minimum height=0.75cm,
        align=center
    },
    lifeline/.style={draw=gray!55, dashed},
    msg/.style={-stealth, thick},
    ret/.style={-stealth, thick, dashed},
    phase/.style={
        draw=gray!40,
        fill=gray!12,
        rounded corners,
        minimum width=14.5cm,
        minimum height=0.55cm,
        font=\bfseries
    },
    fragment/.style={
        draw=gray!45,
        rounded corners=2pt,
        fill=gray!3,
        fill opacity=0.25,
        text opacity=1,
        line width=0.35pt
    },
    fragmentlabel/.style={
        draw=gray!45,
        fill=white,
        fill opacity=0.85,
        text opacity=1,
        rounded corners=1pt,
        font=\scriptsize\bfseries,
        inner xsep=4pt,
        inner ysep=2pt
    },
    fragmentcond/.style={
        font=\scriptsize\itshape,
        anchor=west,
        text opacity=1
    }
]

\node[participant] (TR)  at (0,0)    {Trader};
\node[participant] (OB)  at (3.5,0)  {Position\\Management};
\node[participant] (ORA) at (7.0,0)  {Oracle};
\node[participant] (COL) at (10.5,0) {Collateral\\Vault};
\node[participant] (CH)  at (14.0,0) {Clearing\\House};

\foreach \p in {TR,OB,ORA,COL,CH}
    \draw[lifeline] (\p.south) -- ++(0,-20.2);

\node[phase] at (7,-1.0) {Minting};

\draw[ret]
    (TR|-0,-1.8) --
    node[above, sloped] {\scriptsize 1. Submit mint request}
    (OB|-0,-1.8);

\draw[ret]
    (OB|-0,-2.5) --
    node[above, sloped] {\scriptsize 2. Request reference prices}
    (ORA|-0,-2.5);

\draw[msg]
    (TR|-0,-3.2) --
    node[above, sloped] {\scriptsize 3. Provide collateral}
    (COL|-0,-3.2);

\draw[msg]
    (OB|-0,-3.9) --
    node[above, sloped] {\scriptsize 4. Validate collateralization constraints}
    (CH|-0,-3.9);

\draw[ret]
    (CH|-0,-4.6) --
    node[above, sloped] {\scriptsize 5. Read reference prices}
    (ORA|-0,-4.6);

\draw[ret]
    (CH|-0,-5.3) --
    node[above, sloped] {\scriptsize 6. Check posted collateral}
    (COL|-0,-5.3);

\draw[fragment]
    (0.45,-6.05) rectangle (14.35,-10.15);

\node[fragmentlabel] at (0.9,-6.05) {alt};
\node[fragmentcond] at (1.35,-6.35) {[collateralization constraints satisfied]};

\draw[ret]
    (CH|-0,-7.2) --
    node[above, sloped] {\scriptsize 7. Approve minting}
    (OB|-0,-7.2);

\draw[msg]
    (OB|-0,-7.9) --
    node[above, sloped] {\scriptsize 8. Credit synthetic asset}
    (TR|-0,-7.9);

\draw[gray!40, dashed, line width=0.3pt]
    (0.45,-8.55) -- (14.35,-8.55);

\node[fragmentcond] at (1.35,-8.85) {[otherwise]};

\draw[msg]
    (CH|-0,-9.8) --
    node[above, sloped] {\scriptsize 7'. Reject mint request / request more collateral}
    (TR|-0,-9.8);

\node[phase] at (7,-11.0) {Position monitoring};

\draw[msg]
    (CH|-0,-11.8) --
    node[above, sloped] {\scriptsize 9. Check collateralization ratio}
    (TR|-0,-11.8);

\draw[fragment]
    (0.45,-12.35) rectangle (14.35,-16.95);

\node[fragmentlabel] at (0.9,-12.35) {alt};
\node[fragmentcond] at (1.35,-12.65) {[compliant]};

\draw[msg]
    (CH|-0,-13.7) --
    node[above, sloped] {\scriptsize 10. Keep position open}
    (TR|-0,-13.7);

\draw[gray!40, dashed, line width=0.3pt]
    (0.45,-14.55) -- (14.35,-14.55);

\node[fragmentcond] at (1.35,-14.85) {[under-collateralized]};

\draw[ret]
    (CH|-0,-15.7) --
    node[above, sloped] {\scriptsize 10'. Execute liquidation}
    (COL|-0,-15.7);

\draw[msg]
    (CH|-0,-16.5) --
    node[above, sloped] {\scriptsize 11. Return residual collateral, if any}
    (TR|-0,-16.5);

\node[phase] at (7,-17.8) {Redeeming the synthetic asset};

\draw[msg]
    (TR|-0,-18.6) --
    node[above, sloped] {\scriptsize 12. Submit redemption request}
    (OB|-0,-18.6);

\draw[ret]
    (OB|-0,-19.4) --
    node[above, sloped] {\scriptsize 13. Unlock collateral}
    (COL|-0,-19.4);

\draw[msg]
    (COL|-0,-20.2) --
    node[above, sloped] {\scriptsize 14. Return collateral}
    (TR|-0,-20.2);


\end{tikzpicture}
\end{adjustbox}
\caption{Sequence diagram of the workflow in synthetic protocols. The diagram distinguishes three main phases: minting, position monitoring, and redemption. Dashed arrows indicate informational interactions without direct value transfer, whereas solid arrows denote value transfers or state-changing actions.}
\label{fig:synthetic-position-sequence}
\end{figure}

\paragraph{Minting}
A trader initiates a synthetic position by submitting a mint request to the protocol’s position-management module (Action~1, Fig.~\ref{fig:synthetic-position-sequence}). The protocol then requests the relevant reference prices from the oracle (Action~2), while the trader provides collateral to the collateral vault (Action~3). Before minting can proceed, the protocol validates the applicable collateralization constraints (Action~4), using both the posted collateral and the reference prices as inputs (Actions~5--6). 

If the required collateralization conditions are satisfied, the protocol approves minting and credits the synthetic asset to the trader (Actions~7--8). Otherwise, the mint request is rejected or additional collateral is required (Action~7', alternative branch). Minting determines both the quantity of synthetic units issued and the corresponding debt position, valued at the synthetic asset’s reference price; together, these define the position’s economic exposure.

\paragraph{Position monitoring}
After issuance, the protocol continuously monitors the health of each open position by checking its collateralization ratio (CR), i.e. the value of posted collateral relative to the outstanding synthetic debt, using updated oracle prices (Action~9). As long as the position remains compliant with the protocol’s collateral requirements, it stays open (Action~10). If instead the position becomes under-collateralized, liquidation may be triggered (Action~10', alternative branch), and any residual collateral is returned to the trader when available (Action~11).

More generally, protocols enforce a minimum collateral ratio to preserve solvency. When a position approaches this threshold, the trader may restore safety by adding collateral, partially repaying the debt, burning synthetic units, or combining these actions. We abstract here from implementation-specific details, such as protocol-specific liquidation penalties or auction-based liquidation procedures.

\paragraph{Redemption}
To close a healthy position, the trader submits a redemption request and repays the outstanding debt by burning the corresponding synthetic asset (Action~12). Once the debt has been extinguished, the protocol unlocks the collateral (Action~13) and returns it to the trader (Action~14). At redemption, any remaining profit or loss, net of redemption fees, stability charges, and execution costs, is realized and reflected in the trader’s final proceeds.

\section{Simulation Framework} \label{math_function}

We complement our systematization of DeFi derivative protocols with numerical simulations that approximate how derivative positions evolve under controlled market and protocol conditions. To this end, we develop a Monte Carlo framework in Python that models the lifecycle of derivative contracts under stochastic price dynamics, protocol-specific fees, and liquidation mechanisms. All code is publicly available on our GitHub,\footnote{
\url{https://github.com/LucaPennella/sok-defi-derivatives}

}
providing a reproducible implementation for testing, comparison, and future extensions.

For the perpetual case analyzed in the main text, the simulation environment is aligned with the tuple-based representation introduced in Section~\ref{sec:concept}. This alignment ensures that the analytical and computational components rely on the same formal specifications, thereby preserving consistency between theoretical modeling and numerical experimentation. A detailed mapping between the analytical tuple components and their corresponding implementation is provided in the repository.
The simulation should therefore be interpreted as a controlled executable instantiation of the perpetual component of the framework, rather than as a full protocol emulator. Although the framework can simulate perpetuals, everlasting options, and synthetic assets, the analysis in this paper focuses on perpetual futures, which currently represent the most widely used class of DeFi derivatives in practice. Prototype implementations for the other instrument classes are included in the associated repository.

In the simulations, the underlying asset price \(P_t\) is modeled as a function of time \(t\) through a geometric Brownian motion, a standard stochastic process in derivatives pricing~\parencite{hull2016options,shreve2004stochastic}:
\[
dP_t = \mu P_t dt + \sigma P_t dW_t,
\]
where \(W_t\) is a standard Brownian motion capturing the stochastic component of price evolution, \(\mu\) denotes the expected rate of return, and \(\sigma\) denotes volatility. Each run generates one realization of \(P_t\) over a fixed horizon \(T\). Along each simulated path, we track the evolution of prices, unrealized profit and loss, trader equity, margin ratios, collateral conditions, liquidation events, and terminal realized PnL. Aggregating multiple runs makes it possible to estimate distributional outcomes and risk metrics, including liquidation probability and realized PnL, under different market and protocol configurations.

The perpetual simulation pipeline combines pathwise analysis, batch experimentation, and sensitivity analysis. Single-path simulations trace the joint evolution of price, equity, PnL, and margin ratios while explicitly incorporating trading, borrowing, and funding fees. Batch simulations repeat this procedure across scenarios to estimate outcome distributions over fixed horizons. Sensitivity analyses then quantify how leverage, volatility, and transaction costs affect risk and profitability. In particular, heatmaps describe the bivariate dependence of realized PnL and liquidation probability on volatility and leverage, while tornado diagrams rank the local univariate impact of individual parameters around a baseline configuration.

\begin{figure}[t]
  \centering
  \begin{subfigure}[t]{0.48\textwidth}
    \centering
    \includegraphics[width=\linewidth]{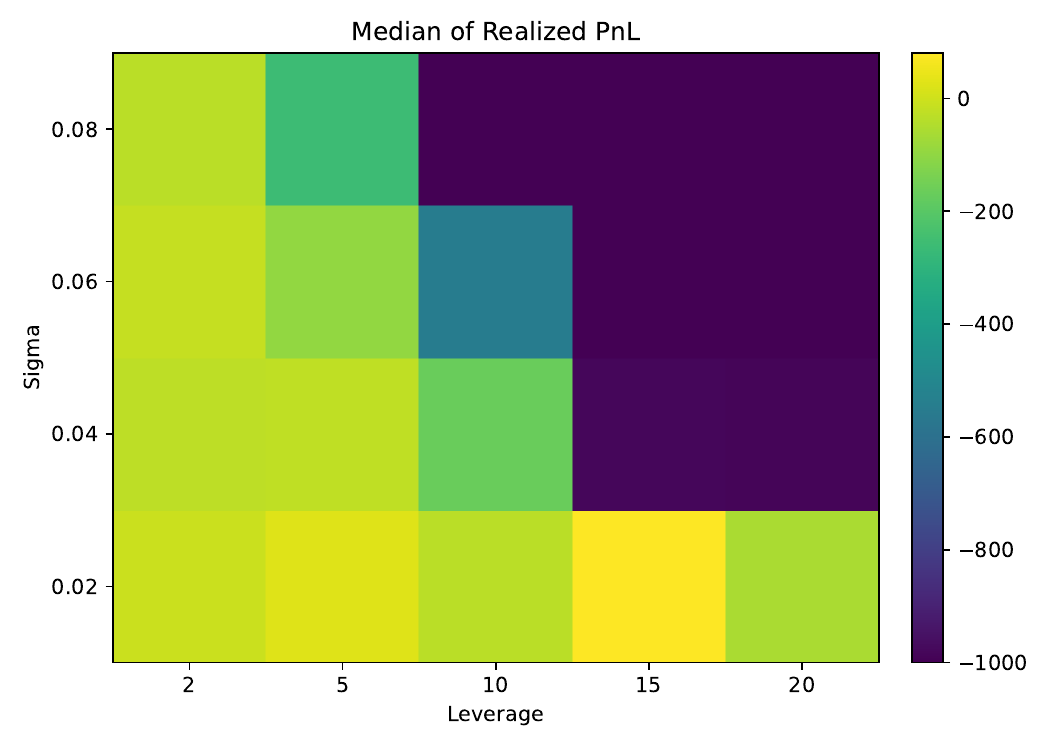}
    \caption{Median realized PnL over $(\sigma,L)$.}
    \label{fig:median_pnl_heatmap}
  \end{subfigure}
  \hfill
  \begin{subfigure}[t]{0.48\textwidth}
    \centering
    \includegraphics[width=\linewidth]{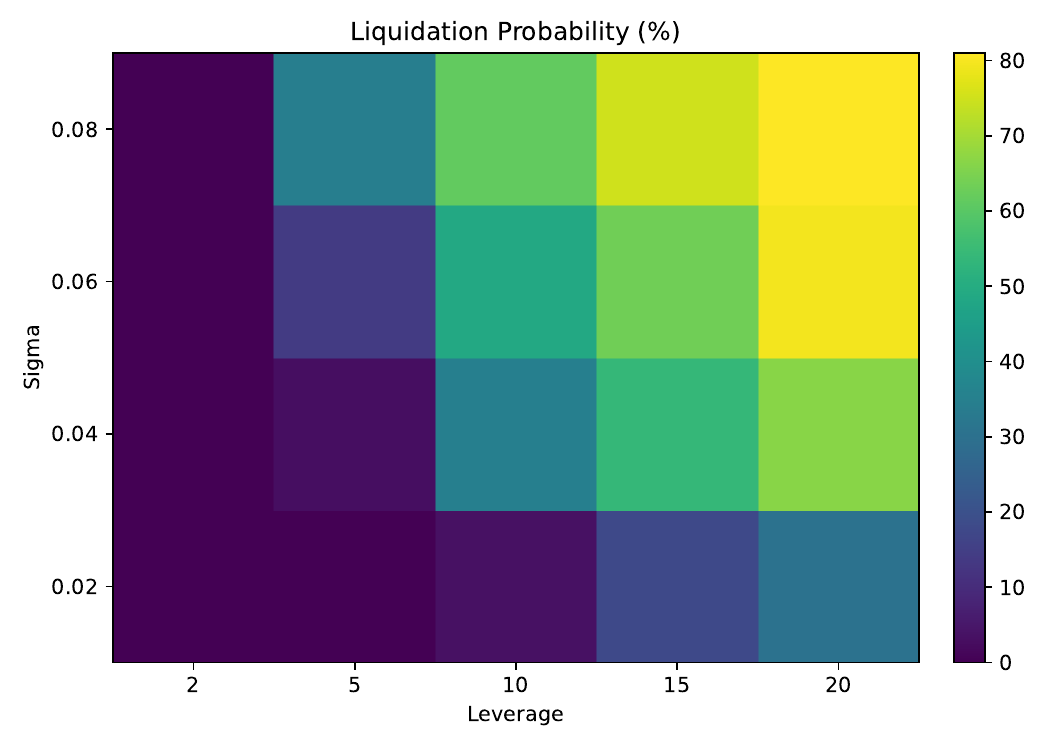}
    \caption{Liquidation probability over $(\sigma,L)$.}
    \label{fig:liq_prob_heatmap}
  \end{subfigure}
  \caption{Sensitivity analysis from Monte Carlo simulations of perpetuals.
  Higher volatility and leverage reduce median PnL and increase liquidation risk.}
  \label{fig:perp_heatmaps}
\end{figure}

To illustrate the behavior of the framework under controlled conditions, we conduct 2000 simulation runs over 7-day horizons across parameter grids spanning volatility values
\(\sigma \in \{0.02, 0.04, 0.06, 0.08\}\)
and leverage levels
\(L \in \{2, 5, 10, 15, 20\}\).
Figure~\ref{fig:perp_heatmaps} summarizes the resulting sensitivity surfaces. Liquidation probability generally rises with both volatility and leverage, suggesting that configurations combining higher market volatility and higher position leverage are associated with more pronounced liquidation risk. Median realized PnL declines as volatility and leverage increase, especially in high-leverage scenarios. Taken together, these results illustrate the structural trade-off embedded in perpetual markets: leverage amplifies exposure, but it also compresses the margin buffer and increases the probability that adverse price movements trigger liquidation.

\begin{figure}[t]
  \centering
  \includegraphics[width=0.78\linewidth]{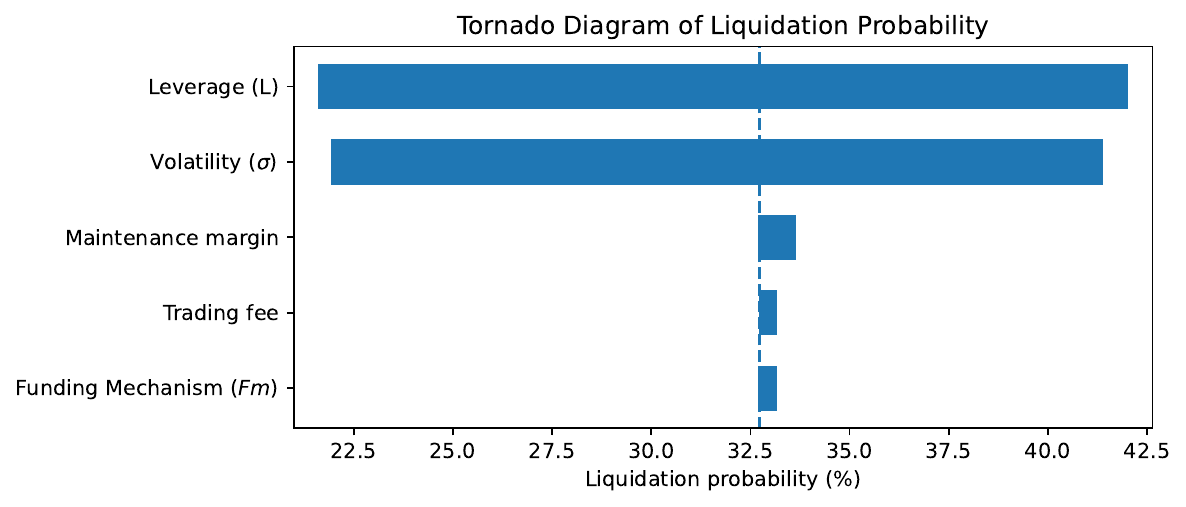}
  \caption{Tornado diagram for 7-day liquidation probability. Bars show the change in liquidation probability under symmetric parameter shocks around the baseline configuration, indicated by the vertical dashed line.}
  \label{fig:tornado_liq}
\end{figure}

We further examine short-term risk drivers through a univariate tornado analysis around the baseline configuration. For each parameter, we apply symmetric \(\pm 20\%\) multiplicative shocks while holding all other parameters fixed, and measure the resulting change in 7-day liquidation probability relative to the baseline value of approximately \(33\%\). As shown in Figure~\ref{fig:tornado_liq}, leverage and volatility are the dominant local drivers of liquidation risk. In this calibration, both parameters generate substantially larger changes in liquidation probability than trading fees, maintenance margin requirements, and funding mechanics, whose local effects remain comparatively small over the 7-day horizon.

These smaller local effects should not be interpreted as evidence that protocol-specific parameters are economically negligible. The tornado diagram measures marginal sensitivity around a specific baseline configuration and over a fixed 7-day horizon; it therefore captures local rather than global importance. Under this calibration, volatility and leverage dominate because they directly affect the distribution of price paths and the size of the margin buffer. By contrast, trading fees, borrowing costs, and funding payments enter the equity process cumulatively and may become more relevant over longer horizons or under alternative fee and funding regimes. Similarly, maintenance margin requirements primarily affect outcomes when simulated paths are close to the liquidation boundary.
Encoding perpetual mechanics through the tuple-based representation makes the formal model executable. The simulation framework therefore operationalizes the analytical representation developed in Section~\ref{sec:concept}, links simulated outcomes to the protocol parameters encoded in \(Z^{\mathrm{prot}}\), and provides a flexible, reproducible tool for studying the risk properties of DeFi derivatives. Appendix~\ref{app:protocol_specific_simulations} reports a protocol-specific implementation of the framework, showing how the general model can be instantiated using parameters inspired by selected perpetual protocols.

\section{Conclusions}
\label{sec:conclusions}
In this paper, we conduct a systematic analysis of derivatives protocols in DeFi.
Although derivative instruments represent a central component of traditional centralized financial markets, their decentralized counterparts remain comparatively understudied, and a unified, high-level formal framework abstracting their design space is currently missing.

This paper addresses this gap by identifying, analyzing, and organizing leading DeFi derivatives protocols into a common analytical framework, focusing on perpetuals, synthetics, and options.

Drawing from protocol documentation, user interfaces, deployed smart contracts, and community-driven sources, we first provide a formal representation of the main classes of decentralized finance derivative instruments. Then we extend the framework to the protocol layer, focusing on the interactions between economic agents and protocol components. The resulting framework captures the main design principles, protocol actors, trading flows, and operational mechanisms that underpin decentralized derivatives protocols. Finally, we conduct numerical simulations showing how derivative positions evolve under varying market and protocol conditions.

By abstracting structural components and financial functions across protocols, our framework provides a basis for systematically comparing heterogeneous implementations.

We acknowledge that our study is not without limitations.
First, our protocol selection process relies on TVL and excludes protocols lacking sufficiently accessible code or documentation. 
This choice may bias the analysis toward relatively mature and well-documented protocols, while overlooking newer, smaller projects that may incorporate innovative mechanisms.
Second, given the early-stage nature of the DeFi derivatives ecosystem, the documentation of some protocols is limited or fragmented. This may limit the comparability of some design features across implementations. 
Third, our analysis reflects a specific observation window and imposes an arbitrary cutoff at December 31 2024, whereas the set of leading protocols is evolving rapidly over time.
Nonetheless, the proposed framework offers a protocol-agnostic abstraction of recurring design principles and provides a structured basis for comparing heterogeneous implementations. It also provides an analytical foundation for studying decentralized derivatives protocols, and serves as a starting point for further theoretical and empirical analyses in the field.
Future work could expand our analysis, framework, and simulation to additional protocols, develop protocol-specific taxonomies for second-layer features, and leverage on-chain data to study user behavior and market efficiency more thoroughly.

As DeFi derivatives continue to evolve, comparative and executable analytical frameworks of this kind can support more systematic research on protocol design, risk formation, and market behavior.

\section*{Acknowledgents}

The contribution of Luca Pennella was funded in part by the PayPal-FNR PEARL Chair in Digital Financial Services, FNR grant reference 13342933/Gilbert Fridgen, and by the FutureFinTech National Centre of Excellence in Research and Innovation, FNR grant reference 16570468, with the support of Luxembourg’s Ministry of Finance. For the purpose of open access, and in fulfilment of the obligations arising from the grant agreements, the author has applied a Creative Commons Attribution 4.0 International (CC BY 4.0) license to any Author Accepted Manuscript version arising from this submission.

\printbibliography

@article{GangwalGT23,
  author       = {Ankit Gangwal and                  Haripriya Ravali Gangavalli and                   Apoorva Thirupathi},
  title        = {A survey of Layer-two blockchain protocols},
  journal      = {J. Netw. Comput. Appl.},
  volume       = {209},
  pages        = {103539},
  year         = {2023},
  OPTurl          = {https://doi.org/10.1016/j.jnca.2022.103539},
  doi          = {10.1016/J.JNCA.2022.103539},
}

@inproceedings{werner2022sok,
author = {Werner, Sam and Perez, Daniel and Gudgeon, Lewis and Klages-Mundt, Ariah and Harz, Dominik and Knottenbelt, William},
title = {SoK: Decentralized Finance (DeFi)},
year = {2023},
publisher = {Association for Computing Machinery},
doi = {10.1145/3558535.3559780},
booktitle = {Proceedings of the 4th ACM Conference on Advances in Financial Technologies},
pages = {30–46},
numpages = {17},
location = {Cambridge, MA, USA},
series = {AFT '22}
}

@book{harvey2021defi,
  title={DeFi and the Future of Finance},
  author={Harvey, Campbell R},
  year={2021},
  publisher={John Wiley \& Sons}
}

@InProceedings{luo2024piercing,
author="Luo, Yichen
and Feng, Yebo
and Xu, Jiahua
and Tasca, Paolo",
editor="Garman, Christina
and Moreno-Sanchez, Pedro",
title="Piercing the Veil of TVL: DeFi Reappraised",
booktitle="Financial Cryptography and Data Security",
year="2026",
publisher="Springer Nature Switzerland",
address="Cham",
pages="3--19",
}

@inproceedings{gudgeon2020defi,
author = {Gudgeon, Lewis and Werner, Sam and Perez, Daniel and Knottenbelt, William J.},
title = {DeFi Protocols for Loanable Funds: Interest Rates, Liquidity and Market Efficiency},
year = {2020},
publisher = {Association for Computing Machinery},
doi = {10.1145/3419614.3423254},
booktitle = {Proceedings of the 2nd ACM Conference on Advances in Financial Technologies},
pages = {92–112},
numpages = {21},
series = {AFT '20}
}

@article{schar:decentralized:2021,
  author  = {Schär, Fabian},
  title   = {Decentralized Finance: On Blockchain- and Smart Contract-Based Financial Markets},
  journal = {Federal Reserve Bank of St. Louis Review},
  year    = {2021},
  volume  = {103},
  number  = {2},
  pages   = {153--174},
  doi     = {10.20955/r.103.153-74}
}

@article{auer2024technology,
  author   = {Auer, Raphael and Haslhofer, Bernhard and Kitzler, Stefan and Saggese, Pietro and Victor, Friedhelm},
  title    = {The Technology of Decentralized Finance ({DeFi})},
  journal  = {Digital Finance},
  year     = {2024},
  month    = mar,
  volume   = {6},
  number   = {1},
  pages    = {55--95},
  issn     = {2524-6186},
  doi      = {10.1007/s42521-023-00088-8}
}

@article{saggese2024assessing,
author = {Pietro Saggese and Esther Segalla and Michael Sigmund and Burkhard Raunig and Felix Zangerl and Bernhard Haslhofer},
title = {Assessing the solvency of virtual asset service providers: are current standards sufficient?},
journal = {Applied Economics},
volume = {57},
number = {49},
pages = {8178--8193},
year = {2025},
publisher = {Routledge},
doi = {10.1080/00036846.2024.2396640}
}

@INPROCEEDINGS{saggese2025towards,
  author={Saggese, Pietro and Fröwis, Michael and Kitzler, Stefan and Haslhofer, Bernhard and Auer, Raphael},
  booktitle={2025 IEEE International Conference on Blockchain and Cryptocurrency (ICBC)}, 
  title={Towards Verifiability of Total Value Locked (TVL) in Decentralized Finance}, 
  year={2025},
  volume={},
  number={},
  pages={1-9},
  doi={10.1109/ICBC64466.2025.11114539}}

@article{xu:sok:2023,
author = {Xu, Jiahua and Paruch, Krzysztof and Cousaert, Simon and Feng, Yebo},
title = {SoK: Decentralized Exchanges (DEX) with Automated Market Maker (AMM) Protocols},
year = {2023},
publisher = {Association for Computing Machinery},
address = {New York, NY, USA},
volume = {55},
number = {11},
doi = {10.1145/3570639},
journal = {ACM Comput. Surv.},
month = feb,
articleno = {238},
numpages = {50},
}

@article{lehar2025decentralized,
author = {Lehar, Alfred and Parlour, Christine},
title = {Decentralized Exchange: The Uniswap Automated Market Maker},
journal = {The Journal of Finance},
volume = {80},
number = {1},
pages = {321-374},
doi = {https://doi.org/10.1111/jofi.13405},
year = {2025}
}

@InProceedings{bartoletti2022theory,
author="Bartoletti, Massimo
and Chiang, James Hsin-yu
and Lluch-Lafuente, Alberto",
editor="Damiani, Ferruccio
and Dardha, Ornela",
title="A Theory of Automated Market Makers in DeFi",
booktitle="Coordination Models and Languages",
year="2021",
publisher="Springer International Publishing",
pages="168--187"
}

@inproceedings{bartoletti:sok:2021,
author = {Bartoletti, Massimo and Chiang, James Hsin-yu and Lafuente, Alberto Lluch},
title = {SoK: Lending Pools in Decentralized Finance},
year = {2021},
publisher = {Springer-Verlag},
doi = {10.1007/978-3-662-63958-0_40},
booktitle = {Financial Cryptography and Data Security. FC 2021 International Workshops: CoDecFin, DeFi, VOTING, and WTSC,  Virtual Event, March 5, 2021,  Revised Selected Papers},
pages = {553–578},
numpages = {26}
}

@article{gogol2024sok,
  title={SoK: liquid staking tokens (LSTs) and emerging trends in restaking},
  author={Gogol, Krzysztof and Velner, Yaron and Kraner, Benjamin and Tessone, Claudio},
  journal={arXiv preprint arXiv:2404.00644},
  year={2024}
}

@article{xu2022reap,
  title={Reap the harvest on blockchain: A survey of yield farming protocols},
  author={Xu, Jiahua and Feng, Yebo},
  journal={IEEE Transactions on Network and Service Management},
  volume={20},
  number={1},
  pages={858--869},
  year={2022},
  publisher={IEEE}
}

@INPROCEEDINGS{cousaert2022sok,
  author={Cousaert, Simon and Xu, Jiahua and Matsui, Toshiko},
  booktitle={2022 IEEE International Conference on Blockchain and Cryptocurrency (ICBC)}, 
  title={SoK: Yield Aggregators in DeFi}, 
  year={2022},
  volume={},
  number={},
  pages={1-14},
  doi={10.1109/ICBC54727.2022.9805523}}

@article{release2023otc,
  title={OTC Derivatives Statistics at end-December 2023},
  author={BIS},
  journal={Bank for International Settlements},
  year={2023}
}

@inproceedings{eskandari2021sok,
author = {Eskandari, Shayan and Salehi, Mehdi and Gu, Wanyun Catherine and Clark, Jeremy},
title = {SoK: oracles from the ground truth to market manipulation},
year = {2021},
publisher = {Association for Computing Machinery},
url = {https://doi.org/10.1145/3479722.3480994},
doi = {10.1145/3479722.3480994},
booktitle = {Proceedings of the 3rd ACM Conference on Advances in Financial Technologies},
pages = {127–141},
numpages = {15},
series = {AFT '21}
}

@book{hull2016options,
  title={Options, futures, and other derivatives},
  author={Hull, John C and Basu, Sankarshan},
  year={2016},
  publisher={Pearson Education India}
}

@misc{defillama,
 key =defillama,
  title        = {Defillama},
  howpublished = {\url{https://defillama.com}},
  note         = {Accessed: January 2025},
year = 2025
}

@online{band:oracle,
	title = {Band Protocol},
	url = {https://bandprotocol.com},
	titleaddon = {Band Protocol},
	urldate = {2024-11-18},
date = {2024-11-18}
}

@online{chainlink:oracle,
	title = {Chainlink: The Foundation for Onchain Finance},
	url = {https://chain.link/},
	shorttitle = {Chainlink},
	urldate = {2024-11-18},
    date = {2024-11-18},
}

@online{jlp,
	title = {{JLP} Economics {\textbar} Jupiter Station},
	url = {https://station.jup.ag/guides/jlp/JLP-Economics},
	urldate = {2024-12-25},
date = {2024-12-25},
	langid = {english},

}

@online{andolfatto:decentralized:2024,
  author   = {Andolfatto, Andrea and Schönleber, Lorenzo},
  title    = {Decentralized and Centralized Options Trading},
  date     = {2024-05-09},
  doi      = {10.2139/ssrn.4822783},
}

@online{perpetual:bitmex,
	title = {Perpetual Contracts Guide - {BitMEX}},
	url = {https://www.bitmex.com/app/perpetualContractsGuide},
	urldate = {2025-02-05},
date = {2025-02-05},
	key="BitMex",
    year="2025"
}

@online{deri:protocol:pricing,
	title = {Pricing Continuously Funded Everlasting Options},
	url = {https://github.com/deri-protocol/whitepaper/blob/master/Pricing_Continuously_Funded_Everlasting_Options.pdf},
	author = {Deri Protocol},
	urldate = {2025-02-07},
date = {2025-02-07}
}

@online{outcomefinance_defillama,
    author = DefiLlama,


	title = {{Outcome Finance DefiLlama}},
	url = {https://defillama.com/protocol/outcome-finance},
	urldate = {2025-03-27},
date = {2025-03-27}
}

@online{opyn,
   author = Opyn,
	title = {Opyn Markets {\textbar} Uniswap, but for perps},
	url = {https://www.opyn.co/},
	urldate = {2025-03-27},
date = {2025-03-27}
}

@online{medium,
	title = {Medium},
	url = {https://medium.com/},
	urldate = {2025-03-27},
date = {2025-03-27}
}

@online{messari,
	title = {Messari},
	url = {https://messari.io/},
	urldate = {2025-03-27},
date = {2025-03-27}
}

@online{dune,
    author = Dune,
	title = {Dune — Crypto Analytics Powered by Community.},
	url = {https://dune.com/home},
	urldate = {2025-03-27},
date = {2025-03-27}
}

@online{aave,
    author = Aave,
	title = {Aave Liquidity Protocol},
	url = {https://app.aave.com/dashboard},
	urldate = {2025-03-29},
date = {2025-03-29},
}

@online{lido,
    author = Lido,
	title = {Lido Liquid Staking},
	url = {https://lido.fi},
urldate = {2025-03-29},
	date = {2025-03-29},

}

@online{uniswap,
key = Uniswap,
	title = {Uniswap Interface},
	url = {https://app.uniswap.org/},
	date = {2025-03-29},
        author = {Uniswap},
        year = {2025}
}

@article{shiller1993measuring,
  author={Shiller, Robert J},
title = {Measuring Asset Values for Cash Settlement in Derivative Markets: Hedonic Repeated Measures Indices and Perpetual Futures},
journal = {The Journal of Finance},
volume = {48},
number = {3},
pages = {911-931},
doi = {https://doi.org/10.1111/j.1540-6261.1993.tb04024.x},
year = {1993}
}

@article{chen2024perpetual,
  author  = {Chen, E. and Ma, M. and Nie, Z.},
  title   = {Perpetual Future Contracts in Centralized and Decentralized Exchanges: Mechanism and Traders' Behavior},
  journal = {Electronic Markets},
  year    = {2024},
  volume  = {34},
  number  = {35},
  doi     = {10.1007/s12525-024-00715-1},
  url     = {https://doi.org/10.1007/s12525-024-00715-1}
}

@article{angeris2023primer,
author = {Angeris, Guillermo and Chitra, Tarun and Evans, Alex and Lorig, Matthew},
title = {Short Communication: A Primer on Perpetuals},
journal = {SIAM Journal on Financial Mathematics},
volume = {14},
number = {1},
pages = {SC17-SC30},
year = {2023},
doi = {10.1137/22M1520931}
}

@InProceedings{do2024novel,
author="Do, Thuat
and Pham, Tuan-Anh
and Tran, Tuan",
editor="Feng, Jun
and He, Songlin
and Zhang, Liang-Jie",
title="Novel Perpetual Futures Market Model Based on a Family of Asymptotic Power Curves",
booktitle="Blockchain -- ICBC 2024",
year="2025",
publisher="Springer Nature Switzerland",
pages="69--83"
}

@article{alexander2020bitmex,
author = {Alexander, Carol and Choi, Jaehyuk and Park, Heungju and Sohn, Sungbin},
title = {BitMEX bitcoin derivatives: Price discovery, informational efficiency, and hedging effectiveness},
journal = {Journal of Futures Markets},
volume = {40},
number = {1},
pages = {23-43},
doi = {https://doi.org/10.1002/fut.22050},
year = {2020}
}

@inproceedings{soska2021towards,
author = {Soska, Kyle and Dong, Jin-Dong and Khodaverdian, Alex and Zetlin-Jones, Ariel and Routledge, Bryan and Christin, Nicolas},
title = {Towards Understanding Cryptocurrency Derivatives:A Case Study of BitMEX},
year = {2021},
publisher = {Association for Computing Machinery},
address = {New York, NY, USA},
doi = {10.1145/3442381.3450059},
booktitle = {Proceedings of the Web Conference 2021},
pages = {45–57},
numpages = {13},
}

@InProceedings{rahimian2024shortfall,
  author =	{Rahimian, Reza and Clark, Jeremy},
  title =	{{A Shortfall in Investor Expectations of Leveraged Tokens}},
  booktitle =	{6th Conference on Advances in Financial Technologies (AFT 2024)},
  pages =	{23:1--23:24},
  series =	{Leibniz International Proceedings in Informatics (LIPIcs)},
  year =	{2024},
  volume =	{316},
  publisher =	{Schloss Dagstuhl -- Leibniz-Zentrum f{\"u}r Informatik},
  address =	{Dagstuhl, Germany},
  doi =		{10.4230/LIPIcs.AFT.2024.23},
}

@article{he2022fundamentals,
  title={Fundamentals of perpetual futures},
  author={He, Songrun and Manela, Asaf and Ross, Omri and von Wachter, Victor},
  journal={arXiv preprint arXiv:2212.06888},
  year={2022}
}

@article{ackerer2024perpetual,
author = {Ackerer, Damien and Hugonnier, Julien and Jermann, Urban},
title = {Perpetual Futures Pricing},
journal = {Mathematical Finance},
volume = {36},
number = {3},
pages = {481-499},
doi = {https://doi.org/10.1111/mafi.70018},
year = {2026}
}

@article{ruan2022perpetual,
  title={Perpetual Futures Contracts and Cryptocurrency Market Microstructure},
  author={Ruan, Qihong and Streltsov, Artem},
  journal={Available at SSRN 4218907},
  year={2022}
}

@INPROCEEDINGS{singh2024option,
  author={Singh, Srisht Fateh and Michalopoulos, Panagiotis and Veneris, Andreas},
  booktitle={2024 IEEE International Conference on Blockchain and Cryptocurrency (ICBC)}, 
  title={Option Contracts in the DeFi Ecosystem: Motivation, Solutions, \& Technical Challenges}, 
  year={2024},
  volume={},
  number={},
  pages={1-7},
  doi={10.1109/ICBC59979.2024.10634452}}

@ARTICLE{al-breiki:trustworthy:2020,
  author={Al-Breiki, Hamda and Rehman, Muhammad Habib Ur and Salah, Khaled and Svetinovic, Davor},
  journal={IEEE Access}, 
  title={Trustworthy Blockchain Oracles: Review, Comparison, and Open Research Challenges}, 
  year={2020},
  volume={8},
  number={},
  pages={85675-85685},
  doi={10.1109/ACCESS.2020.2992698}}

@article{konczal2025pricing,
  author  = {Kończal, Julia},
  title   = {Pricing options on the cryptocurrency futures contracts},
  journal = {International Journal of Theoretical and Applied Finance},
  volume  = {0},
  number  = {0},
  pages   = {2650015},
  year    = {2026},
  doi     = {10.1142/S0219024926500159}
}

@article{brini2024pricing,
title = {Pricing cryptocurrency options with machine learning regression for handling market volatility},
journal = {Economic Modelling},
volume = {136},
pages = {106752},
year = {2024},
doi = {https://doi.org/10.1016/j.econmod.2024.106752},
author = {Alessio Brini and Jimmie Lenz},
}

@article{cheng2021liquidation,
author = {Zhiyong Cheng and Jun Deng and Tianyi Wang and Mei Yu},
title = {Liquidation, leverage and optimal margin in bitcoin futures markets},
journal = {Applied Economics},
volume = {53},
number = {47},
pages = {5415--5428},
year = {2021},
publisher = {Routledge},
doi = {10.1080/00036846.2021.1922597},
}

@ARTICLE{zhang2024practical,
  author={Zhang, Zhao and Xu, Chunxiang and Jiang, Changsong},
  journal={IEEE Transactions on Services Computing}, 
  title={Practical Blockchain-Based Options Contract}, 
  year={2024},
  volume={17},
  number={6},
  pages={4097-4110},
  doi={10.1109/TSC.2024.3404372}}

@article{rahman2022systematization,
  title={Systematization of knowledge: Synthetic assets, derivatives, and on-chain portfolio management},
  author={Rahman, Abrar and Shi, Victor and Ding, Matthew and Choi, Elliot},
  journal={arXiv preprint arXiv:2209.09958},
  year={2022}
}

@article{de2022arbitrage,
author = {De Blasis, Riccardo and Webb, Alexander},
title = {Arbitrage, contract design, and market structure in Bitcoin futures markets},
journal = {Journal of Futures Markets},
volume = {42},
number = {3},
pages = {492-524},
doi = {https://doi.org/10.1002/fut.22305},
year = {2022}
}

@article{alexander2023hedging,
title = {Hedging with automatic liquidation and leverage selection on bitcoin futures},
journal = {European Journal of Operational Research},
volume = {306},
number = {1},
pages = {478-493},
year = {2023},
doi = {https://doi.org/10.1016/j.ejor.2022.07.037},
author = {Carol Alexander and Jun Deng and Bin Zou},
}

@book{shreve2004stochastic,
  title={Stochastic calculus for finance II: Continuous-time models},
  author={Shreve, Steven E and others},
  volume={11},
  year={2004},
  publisher={Springer}
}

@InProceedings{kitzler2024governance,
author="Kitzler, Stefan
and Balietti, Stefano
and Saggese, Pietro
and Haslhofer, Bernhard
and Strohmaier, Markus",
editor="Clark, Jeremy
and Shi, Elaine",
title="The Governance of Decentralized Autonomous Organizations: A Study of Contributors' Influence, Networks, and Shifts in Voting Power",
booktitle="Financial Cryptography and Data Security",
year="2025",
publisher="Springer Nature Switzerland",
pages="313--330"
}

@article{han2025review,
title = {A review of DAO governance: Recent literature and emerging trends},
journal = {Journal of Corporate Finance},
volume = {91},
pages = {102734},
year = {2025},
doi = {https://doi.org/10.1016/j.jcorpfin.2025.102734},
author = {Jungsuk Han and Jongsub Lee and Tao Li}
}

@InProceedings{kraner_et_al:LIPIcs.AFT.2025.9,
  author =	{Kraner, Benjamin and Pennella, Luca and Vallarano, Nicol\`{o} and Tessone, Claudio J.},
  title =	{{Money in Motion: Micro‑Velocity and Usage of Ethereum’s Liquid Staking Tokens}},
  booktitle =	{7th Conference on Advances in Financial Technologies (AFT 2025)},
  pages =	{9:1--9:18},
  series =	{Leibniz International Proceedings in Informatics (LIPIcs)},
  year =	{2025},
  volume =	{354},
  publisher =	{Schloss Dagstuhl -- Leibniz-Zentrum f{\"u}r Informatik},
  address =	{Dagstuhl, Germany},
  doi =		{10.4230/LIPIcs.AFT.2025.9}
}

@misc{vella2026taxonomyrealworldassettokenization,
      title={A Taxonomy of Real-World Asset Tokenization for Blockchain-Based Financial Infrastructure}, 
      author={Giorgio Vella and Luca Pennella and Mark C. Ballandies},
      year={2026},
      archivePrefix={arXiv},
      doi = {10.48550/arXiv.2606.08534}
}

\section{Appendix}
\label{sec:app}

\subsection{Derivatives contracts in traditional finance}

This section provides additional details on the derivative securities traded in traditional financial markets discussed in Section~\ref{sec:background}.  
We outline their main properties, how they are traded, and describe their payoff functions. This serves as a support for readers that are not familiar with such concepts. A deeper discussion can be found in~\parencite{hull2016options}.

\paragraph{Forward contracts} are contracts between two parties to buy or sell an asset at a specified future date $T$ for a price previously agreed upon, the \textit{delivery price} $K$. It differs from the \textit{spot price}, $S_0$, the price one would pay for immediate delivery (i.e., at time $t = 0$) of the asset.
The party agreeing to buy opens a long position, whereas the one agreeing to sell opens a short position. 
The delivery price can differ from the spot price at maturity $S_T$, that is, the actual spot price on the day the contract expires -- and this determines the profitability of the contract for the involved parties. The payoffs of the buyer and seller are respectively given by the formulas:

\[\Pi_b = S_T - K, \qquad \Pi_s = K - S_T\ \]

Figure~\ref{fig:forw} illustrates graphically how the payoff of a long position changes as a function of $S_T$. The payoff of a short position is symmetrical with respect to the x-axis.

Forwards are private, non-standardized, large-size contracts whose terms (e.g., quantity, quality, expiration) can be negotiated directly between the parties. These types of contracts are traded in over-the-counter (OTC) markets, typically by large institutional participants, interacting through a network of brokers or electronic communication networks. In contrast to exchange markets with visible order books and transparent price discovery, in OTC markets, prices are set by large financial institutions that act as market makers, i.e. they provide liquidity by quoting buy and sell prices at which they are ready to accept incoming orders.
Typically, forwards have one specified delivery date and are settled at the end of the contract term.

\begin{figure}[h!]
	\begin{tikzpicture}
		\def\K{3}
		
		\draw[->] (-0.5, 0) -- (6.5, 0) node[below] {\shortstack{Spot Price \\ ( $S_T$ )}};
		\draw[->] (0, -2) -- (0, 2.2) node[left] {\shortstack{Payoff \\ ( $\Pi$ )}};
		
		\node[below] at (\K, 0) {$K$};
		
		\draw[thick, blue, domain=0:6] plot (\x, {(0.5*\x - 0.5*\K)});
		\node[above right, blue] at (4.5, 1.6) {\shortstack{Long Position: \\ $ \Pi = S_T - K$}};
		
	\end{tikzpicture}
	\caption{Payoff diagram with the spot price at expiration ($S_T$) on the x-axis and the payoff on the y-axis.}

	\label{fig:forw}
\end{figure}

\paragraph{Futures} are similar to forward contracts; however, these are standardized and more liquid contracts whose terms are uniform across markets.
Therefore, futures are traded primarily in exchange-traded markets. Buyers and sellers submit their orders to the exchange. Orders are organized into an order-book that sorts them based on price and quantity bid (or asked). Orders are then matched according to specific algorithms. When this occurs, a trade is executed, and orders are cleared from the order-book.
The price discovery is the process by which an asset price is determined through supply and demand in the marketplace.

In terms of their mathematical structure at expiration, forward and futures contracts have identical payoff functions.

The main difference resides in that futures contracts are settled on a daily basis, while forward contracts are settled only at expiration.
Users entering into a future contract are required to deposit an amount known as the initial margin, i.e. a percentage of the contract's total value, to ensure they can fulfill their obligations.
As futures are traded, their value might vary. At the end of each trading day, a trade reflecting such changes is settled, leading to funds moving from long to short positions (or vice versa). This process is called marking-to-market.
Thus, gains or losses are realized incrementally each day as the contract price changes.

\begin{figure*}[t]
  \centering

  \begin{subfigure}[t]{0.48\textwidth}
    \centering
    \begin{tikzpicture}[x=0.7cm, y=0.7cm]
      \def\K{3}

      \draw[->] (-0.2, 0) -- (5.2, 0)
        node[below right] {\shortstack{Spot Price \\ ($S_T$)}};
      \draw[->] (0, -1) -- (0, 2.1)
        node[above left] {\shortstack{Payoff \\ ($\Pi$)}};

      \node[below] at (\K, 0) {$K$};

      \draw[thick, blue, domain=0:\K]   plot (\x, {0});
      \draw[thick, blue, domain=\K:5]   plot (\x, {0.5*\x - 0.5*\K});
      \node[blue, font=\scriptsize, align=center, text width=3cm]
        at (2.0,1.3)
        {Call option holder\\$\max(S_T - K, 0)$};

      \draw[thick, dashed, domain=0:\K] plot (\x, {-0.6});
      \draw[thick, dashed, domain=\K:5] plot (\x, {0.5*\x - 0.5*\K - 0.6});
      \node[anchor=west, font=\scriptsize] at (2.4, -0.9) {Profit};
    \end{tikzpicture}
  \end{subfigure}
  \hfill
  \begin{subfigure}[t]{0.48\textwidth}
    \centering
    \begin{tikzpicture}[x=0.7cm, y=0.7cm]
      \def\K{3}

      \draw[->] (-0.2, 0) -- (5.2, 0)
        node[below right] {\shortstack{Spot Price \\ ($S_T$)}};
      \draw[->] (0, -1) -- (0, 2.1)
        node[above left] {\shortstack{Payoff \\ ($\Pi$)}};

      \node[below] at (\K, 0) {$K$};

      \draw[thick, blue, domain=0:\K]   plot (\x, {0.5*\K - 0.5*\x});
      \draw[thick, blue, domain=\K:5]   plot (\x, {0});
      \node[blue, font=\scriptsize, align=center, text width=3cm]
        at (3.0,1.3)
        {Put option holder\\$\max(K - S_T, 0)$};

      \draw[thick, dashed, domain=0:\K] plot (\x, {0.5*\K - 0.5*\x - 0.6});
      \draw[thick, dashed, domain=\K:5] plot (\x, {-0.6});
      \node[anchor=west, font=\scriptsize] at (2.4, -0.9) {Profit};
    \end{tikzpicture}
  \end{subfigure}

  \caption{Payoff diagram with the spot price at expiration ($S_T$) on the x-axis and the payoff on the y-axis.}
  \label{fig:option}
\end{figure*}

\paragraph{Options} are instruments that grant the possibility, but not the obligation, to exercise a certain right on an underlying asset within a specific period.
The most common ones are call and put options. They respectively allow the option buyer (holder) to purchase or sell the underlying asset at a strike (also called exercise) price $K$. The seller of the option is also called the writer.
American and European options, respectively, refer to contracts that can be exercised at any time or only on the expiration date itself.

In terms of payoff functions, the payoffs of the holder and writer of a call option are respectively given by the formulas:
\[
	\Pi_h = 
	\begin{cases} 
		S_T - K & \text{if } S_T > K \\
		0 & \text{if } S_T \leq K 
	\end{cases}
	\quad \quad
	\Pi_w = 
	\begin{cases} 
		K - S_T & \text{if } S_T > K \\
		0 & \text{if } S_T \leq K 
	\end{cases}
	\]

The payoffs of the holder and writer of a put option are respectively given by the formulas:
\[
	\Pi_h = 
	\begin{cases} 
		0 & \text{if } S_T \geq K \\
		K - S_T & \text{if } S_T < K 
	\end{cases}
	\quad \quad
	\Pi_w = 
	\begin{cases} 
		0 & \text{if } S_T \geq K \\
		S_T - K & \text{if } S_T < K 
	\end{cases}
	\]

Notably, the holder of the option must pay, for this right, a premium to the writer of the option. Therefore, the final profit accounts for the premium paid (received) to buy (sell) the option.
Figure~\ref{fig:option} illustrates graphically how the payoff and profit of a call (left) and put (right) position changes for the holder as a function of $S_T$. The payoff of a short position is symmetrical with respect to the x-axis.
The buyer's maximum loss is limited to the original purchase price of the option contract, while its potential gains are theoretically unlimited if the spot price rises far above the strike price.

\subsection{Data resources} 
Table~\ref{tab:resr} reports the data sources used for this study.
\begin{table}[htbp]
\caption{Data resources on DeFi protocols and documentation}
\label{tab:resr}
\centering
\footnotesize
{
  \setlength{\tabcolsep}{3pt} %

  \begin{tabular*}{0.98\linewidth}{@{\extracolsep{\fill}}l
p{0.23\linewidth}
  p{0.40\linewidth}@{}}
  \toprule
  \textbf{Source} & \textbf{Description} & \textbf{Link} \\
  \midrule
  Alchemix                 & Official Documentation                        & \url{https://docs.alchemix.fi/} \\

  Derive                   & Official Documentation                        & \url{https://docs.derive.xyz/docs/about-derive} \\

  Deri                     & Official Documentation                        & \url{https://docs.deri.io/} \\
  Deri                     & V4 white paper                                & \url{https://github.com/deri-protocol/whitepaper/blob/master/deri_v4_whitepaper.pdf} \\
  Deri                     & V3 white paper                                & \url{https://github.com/deri-protocol/whitepaper/blob/master/deri_v3_whitepaper.pdf} \\
  Deri                     & V2 white paper                                & \url{https://github.com/deri-protocol/whitepaper/blob/master/deri_v2_whitepaper.pdf} \\
  Deri                     & V1 white paper                                & \url{https://github.com/deri-protocol/whitepaper/blob/master/deri_whitepaper.pdf} \\
  Deri                     & white paper Exchange Everlasting Option       & \url{https://github.com/deri-protocol/whitepaper/blob/master/deri_everlasting_options_whitepaper.pdf} \\
  Deri                     & white paper Pricing Everlasting Option with Interest Rate & \url{https://github.com/deri-protocol/whitepaper/blob/master/deri_everlasting_options_with_interst_rate.pdf} \\
  Deri                     & white paper Pricing Everlasting Option        & \url{https://github.com/deri-protocol/whitepaper/blob/master/Pricing_Continuously_Funded_Everlasting_Options.pdf} \\

  Drift        & Official Documentation                        & \url{https://docs.drift.trade/} \\

  Drift        & White Paper                                   & \url{https://cdn.prod.website-files.com/611580035ad59b20437eb024/61293b57e3103934ddc5535f_v0%20Devnet%20Feature%20Paper%20-%20Revision%201.1.pdf} \\

  dYdX                     & Support Documentation                         & \url{https://help.dydx.trade/en/} \\
  dYdX                     & Official Documentation                        & \url{https://docs.dydx.exchange/} \\

  GammaSwap                & Official Documentation                        & \url{https://docs.gammaswap.com/} \\

  GMX                      & Official Documentation                        & \url{https://gmx-docs.io/docs/intro/} \\

  Hegic                    & Official Documentation                        & \url{https://www.hegic.co/app#/learn/} \\
  Hegic                    & White Paper                                   & \url{https://github.com/hegic/whitepaper/blob/master/Hegic%20Protocol%20Whitepaper.pdf} \\

  Hyperliquid              & Official Documentation                        & \url{https://hyperliquid.gitbook.io/hyperliquid-docs} \\

  Jupiter                  & Support Documentation                         & \url{https://support.jup.ag/hc/en-us} \\
  Jupiter                  & Developer Documentation                       & \url{https://dev.jup.ag/} \\

  Metronome                & Official Documentation                        & \url{https://docs.metronome.io/metronome-2.0/master} \\

  Opyn                     & Official Documentation                        & \url{https://opyn.gitbook.io/opyn-hub} \\

  Paradigm                 & White Paper Power Perpetuals                  & \url{https://www.paradigm.xyz/2021/08/power-perpetuals} \\
  Paradigm                 & Everything is a perp                          & \url{https://www.paradigm.xyz/2024/03/everything-is-a-perp} \\
  Paradigm                 & Floor Perps                                   & \url{https://www.paradigm.xyz/2021/08/floor-perps} \\
  Paradigm                 & Everlasting Option                            & \url{https://www.paradigm.xyz/2021/05/everlasting-options} \\

  Perp v2                  & Official Documentation                        & \url{https://support.perp.com/} \\

  Solscan                  & 2° Perp On-chain Transaction                  & \url{https://solscan.io/tx/39aHLoPEEjpsN625BhuGpgYR2Li1nLFgdRZASnrEPvMCQU1swbWR551iTw81uW7h5udaQVwFDzJ2Mh19E8N8Bc8h} \\
  Solscan                  & 1° Perp On-chain Transaction                  & \url{https://solscan.io/tx/4JxAdCXhGsgZeujHuDc7aE8b4NPWuSEegME2T2UppufjWvW8Df7ypzBtmagQTG2qmXCrtYzBjsXETDBrfmJxsGqj} \\

  Synthetix                & Official Documentation                        & \url{https://docs.synthetix.io/} \\

  Taiga                    & Official Documentation                        & \url{https://docs.taigaprotocol.io/} \\

  Youves                   & Official Documentation                        & \url{https://docs.youves.com/} \\
  Youves                   & White Paper                                   & \url{https://docs.youves.com/assets/files/Quantitative_Paper_-_Synthetic_USD_on_Tezos-5148c0a85eda3c82147a59f9995009c5.pdf} \\
  \bottomrule
  \end{tabular*}
}
\end{table}

\subsection{Perpetuals Protocols - Fees Glossary}

\begin{table}[htbp]
\caption{Glossary of technical terms used in perpetual-futures fee tables}\label{tab:gloss}
\centering
\scriptsize
\begin{tabular}{l p{10cm}}
\toprule
\textbf{Term} & \textbf{Meaning} \\
\midrule
Ask price & Lowest price at which a seller is currently willing to sell the asset. \\[3pt]

Bid price & Highest price at which a buyer is currently willing to purchase the asset. \\[3pt]

Index price & Volume-weighted average of spot prices across several exchanges, used as the “fair” reference price for the perpetual contract. \\[3pt]

Mid-price & Simple average of best bid and best ask: \(\tfrac{\text{Bid}+\text{Ask}}{2}\). \\[3pt]

EWMA  & Moving average that gives more weight to recent prices; decay factor is chosen by the protocol (e.g.\ 1-hour window). \\[3pt]

Oracle price & Price delivered by an external data feed (oracle) rather than the protocol’s own order book. \\[3pt]

Funding fee / funding rate & Hourly (or 8-hour, etc.) payment exchanged between long and short traders to keep the perp price aligned with the index price. Positive funding means longs pay shorts; negative means the opposite. \\[3pt]

Borrow fee & Interest charged for borrowing liquidity from the pool (leveraged positions). Computed continuously, settled hourly. \\[3pt]

Utilization ratio & \(\text{Locked liquidity} \div \text{Total liquidity in pool}\); indicates how “full” the lending pool is. \\[3pt]

Clamp function & \(\text{clamp}(x, a, b)\) limits a value \(x\) to the interval \([a, b]\); values below \(a\) return \(a\), above \(b\) return \(b\). Used in Hyperliquid’s funding formula. \\[3pt]

Price-impact fee & Extra spread charged when a trade is large enough to move the pool price beyond a set threshold. \\[3pt]

Maker / taker & Fee schedule where “makers” add liquidity (post orders) and “takers” remove it (hit existing orders); makers usually pay lower fees. \\[3pt]

Premium (funding premium) & Relative difference between the perp market price and the index price; determines who pays the funding fee. \\[3pt]

\bottomrule
\end{tabular}
\end{table}

\subsection{Protocol-Specific Simulation Parameterization}
\label{app:protocol_specific_simulations}

We further adapt the simulation settings to replicate parameters used by specific protocols. For illustration, we simulate a long perpetual position on Solana configured with values inspired by Jupiter Exchange. The simulation spans a 7‑day horizon across grids of leverage \(L \in \{2, 5, 10, 15, 20, 50, 100\}\) and volatility \(\sigma \in \{0.02, 0.04, 0.06, 0.08\}\), each repeated 500 times. Key parameters include: 
\begin{itemize}
    \item initial collateral: 1000.0 units,
    \item trading fee rate: 0.0006 (on open and close),
    \item borrow fee rate: 0.000027 per time step,
    \item maintenance margin rate: 0.002556,
    \item slippage on entry: 20 basis points,
    \item funding fees: excluded in this configuration.
\end{itemize}

The goal is to compute liquidation probabilities under varying levels of leverage and volatility. Table~\ref{tab:jupiter_liq} reports the results. At low volatility (\(\sigma = 0.02\)), liquidation risk remains limited even at moderate leverage, whereas at high volatility (\(\sigma = 0.08\)) it exceeds 75\% for \(L \geq 15\) and approaches certainty at extreme leverage. This nonlinear pattern highlights how risk grows disproportionately with volatility and position size.

\begin{table}[htbp]
\centering
\caption{Liquidation probability (\%) for a long Solana perpetual varying volatility $\sigma$ and leverage $L$.}
\label{tab:jupiter_liq}
\footnotesize
\begin{tabular}{c|ccccccc}
\toprule
$\sigma$ & $L=2$ & $L=5$ & $L=10$ & $L=15$ & $L=20$ & $L=50$ & $L=100$ \\
\midrule
0.02 & 0.0 & 0.0 & 5.6 & 20.2 & 35.0 & 72.8 & 85.4 \\
0.04 & 0.0 & 3.4 & 32.4 & 50.2 & 61.0 & 86.0 & 92.0 \\
0.06 & 0.0 & 16.0 & 53.2 & 65.2 & 75.2 & 88.2 & 94.8 \\
0.08 & 0.0 & 27.4 & 63.0 & 75.2 & 80.8 & 91.0 & 94.8 \\
\bottomrule
\end{tabular}
\end{table}

\end{document}